\documentclass[12pt,preprint]{aastex} 

\usepackage{natbib}
\usepackage{color}
\usepackage{rotating}
\usepackage{lscape}
\usepackage{longtable}

\newcommand{\tfwhm}{{t_\mathrm{FWHM}}}

\newcommand{\rps}{\ensuremath{r_{\rm P1}}}
\newcommand{\ips}{\ensuremath{i_{\rm P1}}}

\begin{document}
%
\title{PANDROMEDA - FIRST RESULTS FROM THE HIGH-CADENCE MONITORING OF M31 WITH PAN-STARRS 1}

\author{C.-H. Lee\altaffilmark{1,2,3}, A. Riffeser\altaffilmark{1,2}, J. Koppenhoefer\altaffilmark{2,1}, S. Seitz\altaffilmark{1,2}, R. Bender\altaffilmark{1,2}, U. Hopp\altaffilmark{1,2}, C. G\"ossl\altaffilmark{1,2}, R. P. Saglia\altaffilmark{2,1}, J. Snigula\altaffilmark{2,1}, W. E. Sweeney\altaffilmark{4}, W. S. Burgett\altaffilmark{4}, K. C. Chambers\altaffilmark{4}, T. Grav\altaffilmark{5}, J. N. Heasley\altaffilmark{4}, K. W. Hodapp\altaffilmark{4}, N. Kaiser\altaffilmark{4}, E. A. Magnier\altaffilmark{4}, J. S. Morgan\altaffilmark{4}, P. A. Price\altaffilmark{6}, C. W. Stubbs\altaffilmark{7}, J. L. Tonry\altaffilmark{4}, R. J. Wainscoat\altaffilmark{4}. }

\altaffiltext{1}{University Observatory Munich, Scheinerstrasse 1, 81679 Munich, Germany}
\altaffiltext{2}{Max Planck Institute for Extraterrestrial Physics, Giessenbachstrasse, 85748 Garching, Germany}
\altaffiltext{3}{Graduate Institute of Astronomy, National Central University, Jhongli 32001, Taiwan}
\altaffiltext{4}{Institute for Astronomy, University of Hawaii at Manoa, Honolulu, HI 96822, USA}
\altaffiltext{5}{Planetary Science Institute, 1700 East Fort Lowell, Suite 106, Tucson, AZ 85719, USA}
\altaffiltext{6}{Department of Astrophysical Sciences, Princeton University, Princeton, NJ 08544, USA}
\altaffiltext{7}{Department of Physics, Harvard University, Cambridge, MA 02138, USA}





%
%
%

\begin{abstract}
The Pan-STARRS 1 (PS1) survey of M31 (PAndromeda) is designed to
identify gravitational microlensing events, caused by bulge and disk
stars (self-lensing) and by compact matter in the halos of M31 and the
Milky Way (halo lensing, or lensing by MACHOs). With the 7 deg$^2$ FOV
of PS1, the entire disk of M31 can be imaged with one single
pointing. Our aim is to monitor M31 with this wide FOV with daily
sampling (20 mins/day).  In the 2010 season we acquired in total 91
nights toward M31, with 90 nights in the $\rps$ and 66 nights in the
$\ips$.  The total integration time in $\rps$ and $\ips$ are 70740s
and 36180s, respectively. As a preliminary analysis, we study a
$40' \times 40'$ sub-field in the central region of M31, a
$ 20' \times 20'$ sub-field in the disk of M31 and a $20' \times 20'$
sub-field for the investigation of astrometric precision. We
demonstrate that the PSF is good enough to detect microlensing events.
We present light curves for 6 candidate microlensing events. This is a
competitive rate compared to previous M31 microlensing surveys.  We
finally also present one example light curve for Cepheids, novae and
eclipsing binaries in these sub-fields.
\end{abstract}


\section{Introduction}
Since the seminal paper by \cite{1986ApJ...304....1P}, microlensing
has been used as an elegant way to probe the massive compact halo
object (MACHO) content in our Milky Way. Various experiments have been
conducted (e.g. MACHO, EROS, OGLE, MOA) targeting the Magellanic Clouds
to constrain the MACHO mass fraction in the Milky Way halo. However,
the results remain uncertain because of the self-lensing contamination
\citep{1994Natur.370..275S,2000ApJ...542..281A,2007A&A...469..387T,2009MNRAS.397.1228W,2010MNRAS.407..189W,2011MNRAS.413..493W,2011arXiv1106.2925W,2011arXiv1105.4615C}.

On the other hand, \cite{1992ApJ...399L..43C} proposed to target M31
to search for MACHOs as dark matter candidates in spiral galaxies.
This is because we can have multiple lines of sight toward M31 to
better discriminate self-lensing from halo lensing. For example, the
far-side of M31 experiences a higher column density of MACHOs than the
near-side, thus a spatial asymmetry of the lensing events should show
up if a significant MACHO component exists in the halo of M31.  A
description of the current status of M31 microlensing searches and the
theory of microlensing in crowded fields is given in
\cite{2010GReGr..42.2101C} and references therein. To summarize this
review, previous M31 projects (e.g. POINT-AGAPE
\citep{2001ApJ...553L.137A}, MEGA \citep{2001ASPC..237..243C}) have
monitored $\approx$ $30' \times 30'$ large fields of M31 on the near and
far side of the disk while avoiding the central bulge. The WeCAPP
\citep{2001A+A...379..362R} team took another approach and monitored a
$17.2' \times 17.2'$ field centered on the M31 center.  They have
detected $\sim$ 10 events in their 11-year campaign (Riffeser et
al. in prep.)

The result of the POINT-AGAPE team suggests the existence of MACHOs 
to explain some of the events \citep{2005A&A...443..911C}, 
while the MEGA team claims that the result of their 14 microlensing 
events \citep{2006A&A...446..855D},
though 2 of them are more likely to be supernovae
\citep{2005ApJ...633L.105C}, is reconcilable with self-lensing
scenario. Few possible issues on the MEGA results have been addressed by the 
analysis of \cite{2006A&A...445..375I, 2007A&A...462..895I}. 
One aspect is that the dust in the M31
disk \citep{2009A&A...507..283M} yields an asymmetric detection
efficiency of true variables and microlensing events and thus has to
be included in a quantitative modeling. The other aspect is that the
total number of events found is still not
large enough to allow for small errors in statistical analyses on the MACHO
fraction of M31.
The investigation of the asymmetric signal, the spatial distribution and
the total number of events requires large fields to be
observed. Another point is the time sampling: Because the
full-width-half-maximum (FWHM) timescale of the microlensing events
in M31 are estimated to be shorter than a few days
\citep{2006ApJS..163..225R}, one requires a higher time-resolution to
detect these events. The POINT-AGAPE and MEGA project have
observations every four nights on average \citep{2003A&A...405...15P}.
\cite{2003ApJ...599L..17R} focused on the $17.2' \times 17.2'$ FOV in
the bulge of M31 with daily observations.  

Ongoing campaigns attempt to monitor the inner region of M31
with higher cadence. For example, the ANGSTROM (Andromeda Galaxy
Stellar Robotic Microlensing) collaboration employs a world-wide 
newtork of 2-meter telescopes to allow for 24-hour survillance of
the inner bulge ($4.6' \times 4.6'$ FOV) of M31 \citep{2006MNRAS.365.1099K}. 
The PLAN (Pixel Lensing Andromeda) collaboration aims for 
consecutive-nights campaigns for two fields (with $13' \times 12.6'$
FOV each) around the inner region of M31 \citep{2009ApJ...695..442C}, which lead to the 
first finite-source microlensing event, OAB-N2, toward M31\citep{2010ApJ...717..987C}. 
The analysis on this event, as well as the bright event GL1/PA-S3\citep{2008ApJ...684.1093R}
and PA-N1\citep{2001ApJ...553L.137A}, favors a MACHO senario relative to self-lensing.  
However, a comprehensive study of the MACHO content in the halo of M31 and constraints on the
self-lensing rate requires a large FOV survey with daily
sampling. This is only possible with the advent of PAndromeda.

PAndromeda is one of the 12 key projects within the scope of
Pan-STARRS 1 \citep[see][for a detailed description of the PS1 system,
optical design and the
imager]{2010SPIE.7733E..12K,2004SPIE.5489..667H,2009amos.confE..40T}.
PAndromeda is designed to identify microlensing events toward M31
with high cadence observations (0.5 h per night for a time span of 5
months per year). With the large FOV of Pan-STARRS 1, it is possible
to monitor the entire disk of M31 with one single exposure (see
Fig. \ref{fig.gpc}).  This enables us to compare the self-lensing
event rate (caused by bulge and disk stars within M31) to the
prediction from the theoretical calculation
\citep{2006ApJS..163..225R} with a certain detection efficiency (which
is given from the set-up of PAndromeda and simulations).  The merit of
PAndromeda is the ability to perform a differential measurement across
the entire M31 disk. With similar PSF and exact cadence for each image
and filter, the detection efficiency mostly varies due to M31
background light.  The impacts of PSF variations and cadence are
rather small.  A discrepancy between the observed and theoretical
self-lensing event rate can indicate that the assumptions in the M31
model under consideration (e.g. stellar population) might need to be
improved \citep{2010A&A...509A..61S}.
 
The microlensing event rate will help us to constrain the
mass-fraction of compact objects in M31 and the Milky Way halo, as
well as the mass function at the low mass end of the M31 bulge.
PAndromeda further aims to shed light on the stellar population
properties of M31 based on the color profiles, surface-brightness
(SFB) fluctuations, resolved stars and variables. It will thus improve
our understanding of the mix of stellar ages and metallicities in the
halo, bulge, stellar streams and dwarfs of M31. This information is
required for an accurate interpretation of the microlensing
events.  For long timescale events, PAndromeda has the potential to
detect sub-stellar objects by means of excursions from the standard
light curve due to companion objects \citep[similar to the candidate
discussed by][]{2004ApJ...601..845A,2009MNRAS.399..219I}. 
In addition to microlensing, the high-cadence of PAndromeda data-set
can also shed light on the variables \citep[see e.g.][]{2004MNRAS.351.1071A,2006A&A...445..423F}
and nova \citep[see e.g.][]{2004MNRAS.353..571D,2006MNRAS.369..257D,2011arXiv1109.6573L} in M31.

This paper is organized as follows. In section \ref{sec.survey} we
present the survey parameters and observations done in 2010.  We give
a detailed description of our data reduction, data analysis and
demonstrate the data quality of the 2010 observation season in section
\ref{sec.data_red}. The variables detected in the PAndromeda data-set
are presented in section \ref{sec.variable}.  An overview of the
microlensing events is shown in section \ref{sec.ML}, followed by an
outlook in section \ref{sec.outlook}.  We conclude the paper in
section \ref{sec.conclusion}.

\section{PAndromeda Survey Parameters and Observations in 2010}
\label{sec.survey}

The PAndromeda project is one of the 12 key projects of PS1. It
monitors M31, the Andromeda galaxy.  The PAndromeda survey is also one
of the Pan-STARRS medium deep fields (MDFs) and is called MD0
alternatively.  PS1 observations are carried out with the 1.8m
Panoramic Survey Telescope and Rapid Response System (Pan-STARRS)
located at Haleakala in Hawaii. The camera used is currently the
largest one in the world.

It consists of 60 detectors in an 8$\times$8 array with empty corners.
Each detector is segmented in an 8$\times$8 array of 590$\times$598
pixel cells with gaps between cells of 19 and 13 pixel widths in the
column and row dimensions, respectively. In addition, between each
detector there are gaps of approximately 300 pixels in both
directions. A single detector therefore consists of a 4872$\times$4824
pixel array. Since pixels have a size of 10 $\mu$m which corresponds
to $0.258''$ in the focal plane the FOV of each detector is
$20.95' \times 20.74'$ and the total FOV of the 60 detector array is
about 7 degrees$^2$.  The layout of the detector surface is
illustrated below in Fig \ref{fig.gpc}.

\begin{figure}
  \centering
  \includegraphics[scale=1]{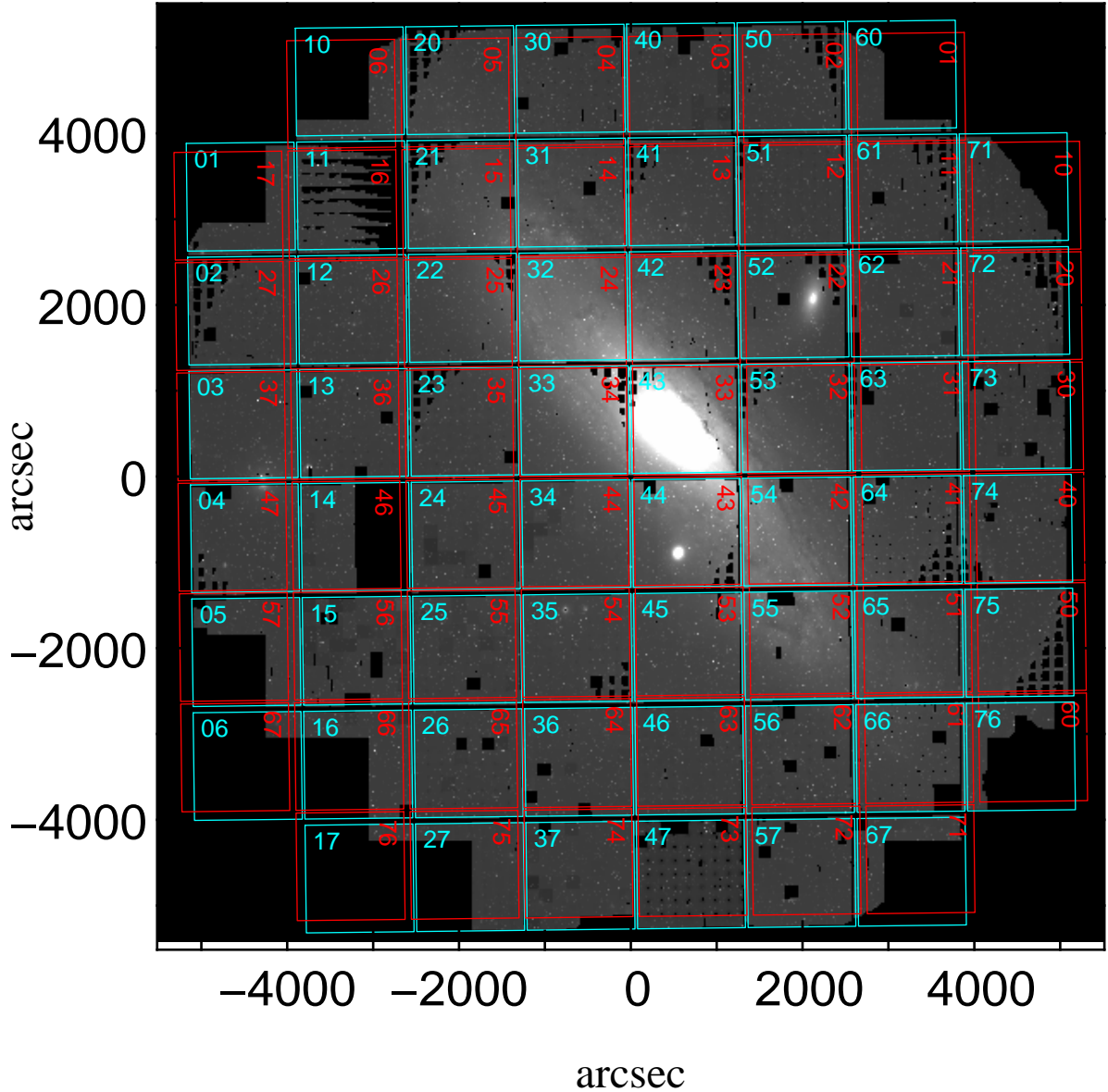}
  \caption{The Giga Pixel Camera (GPC) and its 60 detectors labeled by the blue and red grids. The red numbering indicates the same detector corresponding to the blue numbering, but rotated with 90 degrees during the observation. 
Each detector is composed of 4872$\times$4824 pixel with $0.258''$
per pixel \citep{2009amos.confE..40T}.  The field of view of GPC is
$\sim$ 7 degree$^2$. The entire Andromeda galaxy, M32 and NGC 205 can
be imaged with one pointing (as shown in the figure). The underlying
M31 image is a single $\rps$-band frame, which was chosen to have
relatively small masked area. The masked areas (in black) visible e.g.  in
the detector 14 (marked in blue) are due to ill-functioning orthogonal
transfer arrays (OTAs), video guide star data and areas on the focal
plane with sub-standard imaging performance  within this
detector. These gaps can only be closed by large dithering or
by rotating the camera.}
  \label{fig.gpc}
\end{figure}

PAndromeda started the first observation season in 2010 and monitored
M31 from 23/07/2010 to 27/12/2010.  The M31 observations are done in
$\rps$ and $\ips$-band.  We use two filters to confirm microlensing
events from their achromaticity.  We acquired 90 nights in $\rps$ and
66 nights in $\ips$. The total number of images and integration times
obtained in the first observing season of PAndromeda are listed in
Table \ref{tab.PAndromeda}.

\begin{table}[!h]
\centering
\caption{PAndromeda integration times in $\rps$- and $\ips$-bands for the 2010 season}
\begin{tabular}{cccc}
\hline
Filter     & Nights$^*$ & Images & Total integration time \\
\hline
$\rps$ & 90(74) & 1179 & 70740s \\
\hline
$\ips$ & 66(56) & 603 & 36180s \\
\hline
\multicolumn{4}{l}{$^*$The number of nights with 2 visits are indicated}\\
\multicolumn{4}{l}{in the parenthesis.}
\end{tabular}
\label{tab.PAndromeda}
\end{table}

\begin{figure}[!ht]
  \centering
  \includegraphics[scale=0.8]{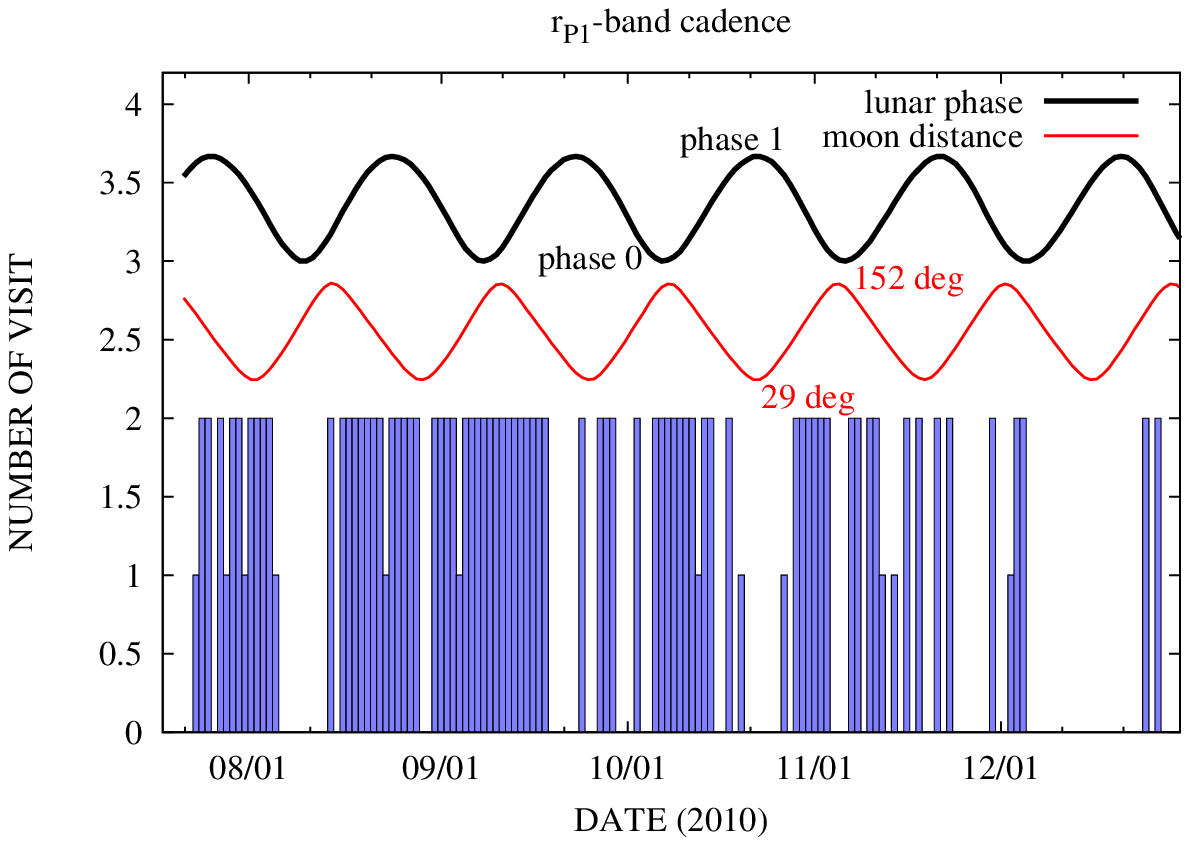}
  \includegraphics[scale=0.8]{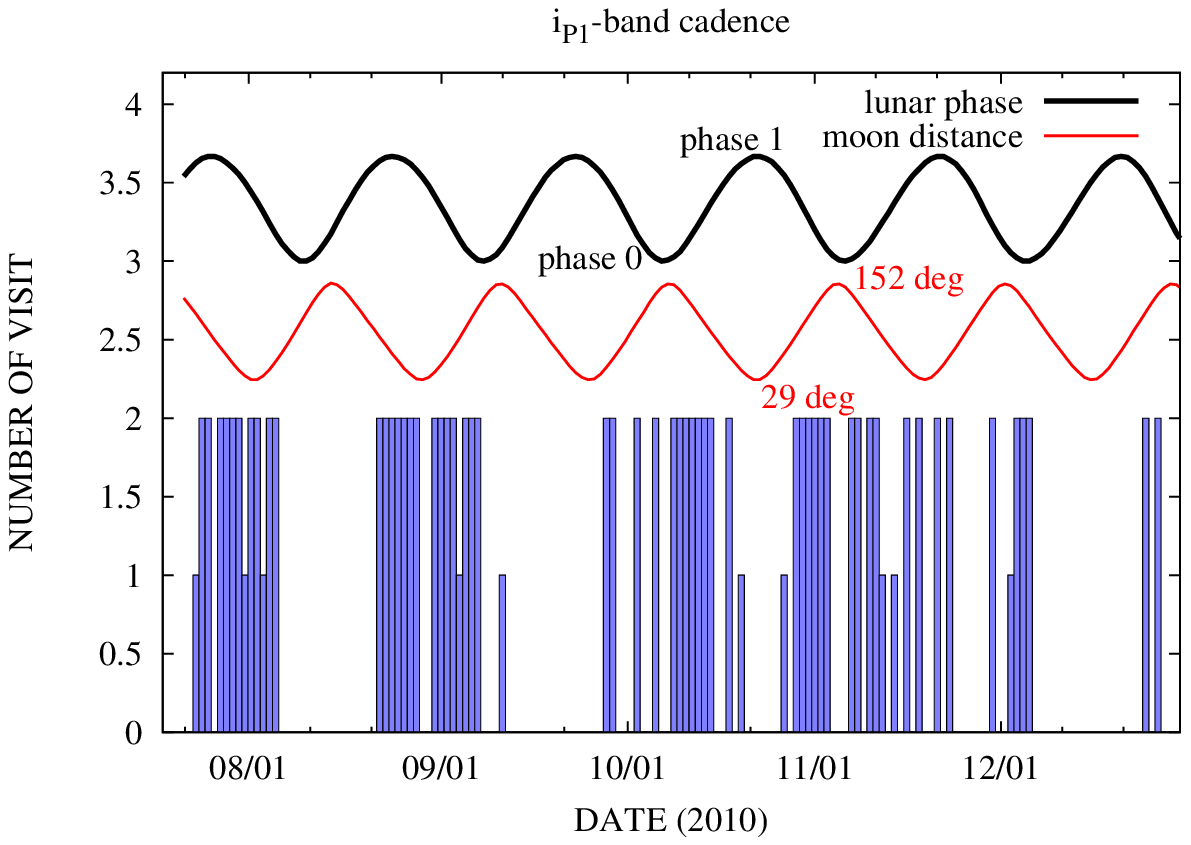}
  \caption{PAndromeda observation cadence. The histogram shows the number of visits 
    (0, 1 or 2 times per night) on M31. The black line indicates the lunar phase (1 for full moon and 0 for 
    new moon). The red line shows the distance of the M31 center from the moon as a function 
    of observing date. A pause in the observations is visible in December when we were reaching 
    the limit of the 2\% PS1-time.}
  \label{fig.cadence}
\end{figure}

The 0.5 h integration time per night was divided into two observation
blocks separated by 3 to 5 hours in order to trace microlensing events
with timescales shorter than 1 day.  The observation cadence is shown
in Fig. \ref{fig.cadence}.

From the data header information one can see that each of the two
nightly observing blocks yielding 12 frames with 60 seconds of
exposure in 2 filters needs 16.2 minutes of telescope time.  The
overhead of about 35 percent includes the 13 seconds of read-out time,
15 seconds for filter change and about 30 seconds for focusing after a
filter change.  Due to the overhead for filter changes we requested to
take only data in the $\rps$ and $\ips$ filters in 2010.  The $\rps$
and $\ips$ filters are chosen because the source of the microlensing
events are expected to be red evolved stars and the contamination from
intrinsic variables in M31 is less severe in the $\rps$ filter.

Note that the central part of the camera close to the bulge of M31 is
slightly out of focus (see Fig. \ref{fig.gpc}).

For each filter, the observations within one visit have a polygonal dithering pattern with a radius of $9''$. It is either a hexagon ($\rps$-band) or a square ($\ips$-band) with the pointings located at the edges and the center.

\section{Data processing}
\label{sec.data_red}
All images obtained for the PAndromeda project are processed through
the Image Processing Pipeline (IPP) \citep{2006amos.confE..50M}. The
IPP runs the images through a succession of stages, including bias and
dark correction, masking and artifact removal, flat-fielding,
astrometric calibration \citep{2008IAUS..248..553M} and a
flux-conserving warping (the images are resampled to a common pixelscale) to a sky-based image plane, so-called
skycells. In the following we use the term ``image'' for the warped
skycells.

The warped images have a dimension of 6000$\times$6000 pixels and a
plate scale of $0.2''$/pixel which is smaller than the natural
plate scale of $0.258''$/pixel of the PS1 Giga Pixel Camera. 
Each image has an attached variance map and a 16-bit mask image.  
For the PAndromeda project all images provided by the IPP are
further processed using the PAndromeda data reduction pipeline. This
pipeline is based on the difference imaging technique
\citep{1998ApJ...503..325A,2002A&A...381.1095G} which provides high
precision photometric measurements in highly crowded fields.  The
PAndromeda pipeline has been integrated by us into the AstroWISE data
management system \citep{2007ASPC..376..491V} which has been developed
to serve surveys with large data volumes. For a detailed description
of the AstroWISE integration of our software we refer to
\cite{AW_Planets}.

\begin{figure}[htbp]
  \begin{center}
    \includegraphics[scale=0.8]{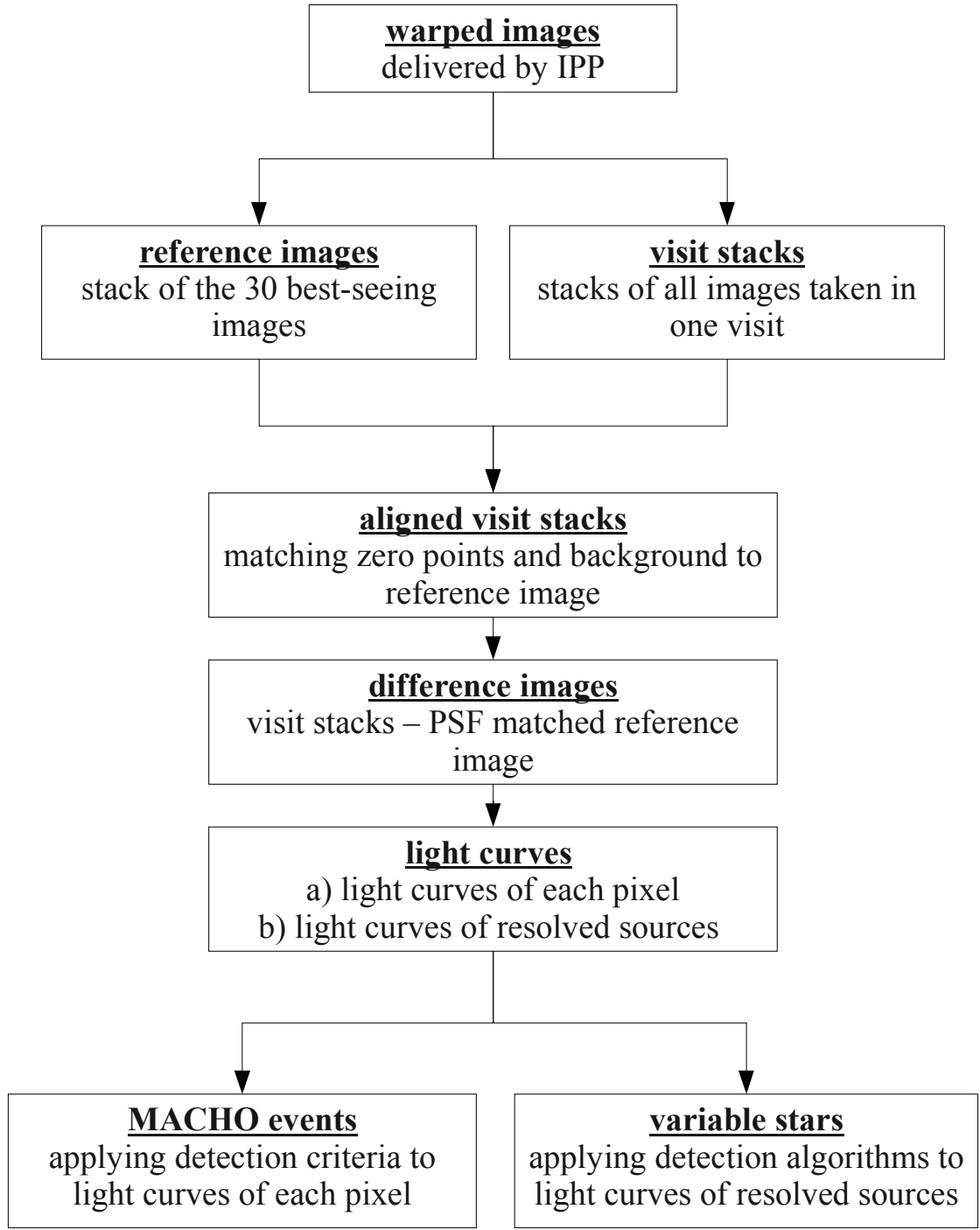}
    \caption{Overview of the processing steps performed by the PAndromeda
      data reduction pipeline for each skycell.}
    \label{fig.flowchart}
  \end{center}
\end{figure}

In the following we describe each processing step that is performed by
the PAndromeda pipeline starting with the warped images until the
creation of the light curves and the detection of microlensing
events. Fig.  \ref{fig.flowchart} gives an overview of the different
processing steps. These steps are performed independently for each
filter $\rps$ and $\ips$ and for each skycell.

At the end of this section we analyze the astrometric and photometric
stability of the light curves (see section \ref{sec.astrometry} and
section \ref{sec.photometry}).\\

\subsection{Interface between IPP and the PAndromeda Pipeline}
\label{sec.pre-processing}
By default the IPP subtracts the background in each warped image using
a bi-linear interpolation of a model with coarsely sampled super
pixels.  The procedure works well in case of pure sky background. In
the PAndromeda survey, however, the images contain also astrophysical
background, i.e. the brightness profile of M31 which is one of the
largest objects on the sky. In the PAndromeda pipeline the brightness
profile is actually used later on. 
Therefore we requested to restore the background by the IPP
that originally had been subtracted.

In the first data release the skycell layout used for the processing
of the PAndromeda data was not optimized. In particular it does not
make use of the fact that the detector boundaries are more or less the
same with respect to M31 in each image.  In some months from now a
reprocessing of the PAndromeda data will be done by the IPP with
skycells that are aligned with the outlines of the detectors.

Although not yet optimal, the processing used for this work is still
good enough for the detection of microlensing events. In total we
obtained 70 skycells with acceptable filling factor.  We require at
least 50\% of each image not to be empty, to save disk space and
computation time by avoiding images with large empty areas.  One
example, skycell 077, is shown in Fig. \ref{fig.ref}. Each skycell
contains pixels of 4 different detectors in the current layout. In
order to save disk space and to speed-up processing we apply a
2$\times$2 pixel binning to all images. In this way we are able to
produce the light curves of a significant fraction of the total FOV of
the first season data within a few weeks of computing time.  In the
future (i.e. when all previously taken PS1 images will be reprocessed
with an improved Image Processing Pipeline) we will switch to the
natural plate scale of $0.258''$/pixel.

In the first processing step of the PAndromeda pipeline all images are
registered to the AstroWISE system, a procedure during which the IPP
variance map and IPP mask images are merged into a single file - a
variance map with zero values for masked pixels.  The IPP variance
maps have correlated noise. Due to remapping and reduced
pixel size this correlated noise is not accounted for in our pipeline,
but we empirically correct for this by rescaling the error bars of the
PSF light curves. (see section \ref{sec.calibration}).

The PSF FWHM of each image is determined using the median of all
FWHM\_IMAGE values measured by SExtractor
\citep{1996A&AS..117..393B}. In addition we apply a filtering
algorithm \citep{2002A&A...381.1095G} to detect and mask cosmic rays
based on their narrow FWHM.

\subsection{Creation of a Reference Image}
\label{sec.refimage}

The difference imaging technique requires the creation of a reference
frame which is a stack of the images with the best image quality. In
our case we select the 30 best seeing images for each skycell
excluding images with a high masking fraction and/or a low
transparency. We create a reference image in $\rps$-band and
$\ips$-band for each skycell. The individual processing steps are as
follows:
\\
In the first step we photometrically align each of the 30 input images
$I_i$ to the image with the very best seeing $I_1$. To account for
zeropoint differences we multiply each image with a scaling factor
$f_i$. Furthermore we add a model $B_i$ to account for differences in
the sky background:
\begin{equation}
	I'_i = f_i \times I_i + B_i \quad .
	\label{eq.scale}
\end{equation}

For a proper photometric alignment of the single images, it is very
difficult to calculate the scaling factor using a set of reference stars. In
order to make use of all pixels in the field of view and to deal with the
high crowding, each scaling factor $f_i$ is calculated with the difference
imaging technique: using all unmasked pixels of the image we calculate
an un-normalized convolution kernel that would match the PSF of image
$I_1$ to image $I_i$ and would simultaneously scale image $I_1$ to
the flux level as image $I_i$. We do not use the kernel to really
convolve the images but rather calculate its inverse norm which
delivers us the scaling factor $f_i$. In this way all scaled images
$I'_i$ have the same zero point but the PSF remains unchanged. Note
that because the sky background in image $I_i$ is still differing from
the sky background in image $I_1$ at this stage, we need to account
for a sky background change while doing the calculation of the
kernel. This is done with including a 1st order polynomial background
in the minimization procedure.

The polynomial background is very robust when calculating the scaling
factor $f_i$, however, for doing a proper adjustment of the sky
background we do not use it and calculate a bicubic spline model $B_i$
which has 16$\times$16 nodes with a separation of 183 pixel. The
spline values at each grid point are calculated in the following way:
first we apply a 61$\times$61 binning to the scaled images $f_i$
$\times$ $I_i$ (excluding all masked pixels) and subtract the binned
images from the binned version of $I_1$. The subtracted images have a
dimension of 49$\times$49 pixel and are then convolved with a
3$\times$3 median kernel resulting in an array of 16$\times$16. These
values are taken as the spline values at the grid points which are
used to calculate the sky background correction $B_i$. After the
photometric alignment all images have the same zeropoint and sky
background.

Recently, \cite{2010MNRAS.409..247K} found that image subtraction in M31 works better 
when first performing a careful photometric alignment prior to matching the 
PSF. Although developed completely independent, our procedure is very 
similar. The difference of our method with respect to \cite{2010MNRAS.409..247K} is that 
we account for PSF differences when doing the photometric alignment.

\begin{figure}[htbp]
  \begin{center}
    \includegraphics[scale=0.8]{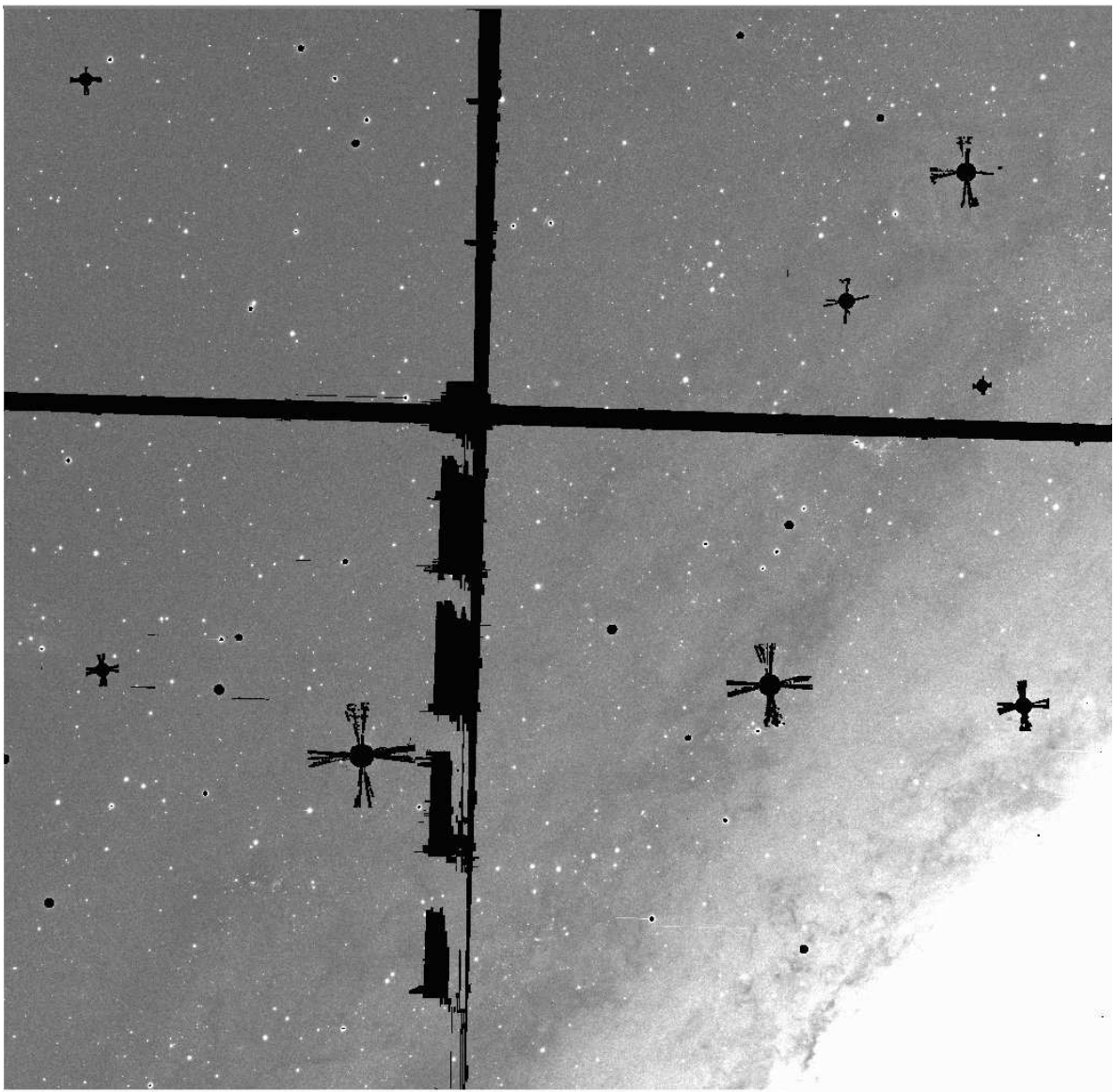}
    \caption{Reference image of skycell 077 in $\rps$-band. The
      skycell contains pixels from 4 adjacent detectors with the vertical
      and horizontal masked area being the gap between them. Note that
      the layout of the skycells has not been optimized for the
      PAndromeda survey. In the future we will use skycells that are
      aligned with the detector borders. Each skycell image 
        has a size of $20' \times 20'$.}
    \label{fig.ref}
  \end{center}
\end{figure}

The reference image is a weighted stack of all 30 photometrically
aligned images where the weight, $\alpha_i$, for each image is
calculated based on the background noise (median of $\sigma$ in each
pixel) and the PSF FWHM, as measured by SExtractor during the image
registration:
\begin{equation}
	\alpha_i = 1 / ( \sigma_i \times \mathrm{FWHM}_i )^2 \quad .
	\label{eq.weights}
\end{equation}
Before stacking we replace masked pixels in each input
image with the pixel values of the other image which has the most
similar PSF. The similarity of the PSF is determined using a set of
isolated stars. If a pixel is masked also in the most similar image,
we replace with the second most similar image and so on. The procedure
to replace masked pixels results in a very homogeneous PSF in the
stacked image which is essential for the following processing
steps. Note that this method only works if the set of input images
already has similar PSFs. Fig. \ref{fig.ref} shows an example
reference image of skycell 077 in $\rps$-band. The median PSF FWHM
of all 30 input images is $0.861''$ with RMS = $0.012''$.
In order to obtain the reference fluxes of all resolved stars in the
reference image we subtract a bicubic spline model and perform PSF
photometry on each source in the background subtracted image. The
source positions are taken from a SExtractor table that has been
created with the parameters DETECT\_THRESH=3 and DETECT\_MINAREA=5. We
iteratively subtract neighboring stars to remove blending effects
\citep[similar to the program DAOphot,][]{1987PASP...99..191S}. The
eference fluxes will be added at a later stage to the fluxes measured
in the difference images in order to create the light curves of
resolved stars (see section \ref{sec.lightcurves}).

\subsection{Creation of a stack for each visit}
\label{sec.nightlystacks} Each visit of M31 consists of (up to) 7
individual exposures in $\rps$-band which are obtained in a dithered
mode. For some visits we also obtained (up to) 5 individual exposures
in $\ips$-band (see section \ref{sec.survey}). The single exposures of
each visit are combined to a visit stack for each filter in order to
improve the S/N and to eliminate masked areas. The procedure to create
visit stacks is identical to the procedure described in the previous
subsection. Also here, masked pixels in one image are replaced by the
pixel values of the image with the most similar PSF. Note that
although the number of input images is lower than in case of the
creation of the reference images the requirement of finding an image
with similar PSF is satisfied since all images of one visit are taken
subsequently after focusing the camera in the beginning of the visit.

\subsection{Photometric alignment}
\label{sec.photalignment}

The single visit stacks are photometrically aligned to the reference
image using the same procedure we described in section
\ref{sec.refimage}. After this step the visit stacks and the reference
image have the same zeropoint and sky background and differ only in
depth and PSF. Note that this zeropoint has an arbitrary value and
only relative fluxes are calibrated. An absolute photometric
calibration is done \textit{a posteriori} as described in section
\ref{sec.lightcurves}.

\subsection{PSF alignment and image subtraction}
\label{sec.PSFalignment}

For each single visit stack we subtract a PSF aligned
reference image and obtain a difference image. The PSF alignment is
done using the standard approach proposed by \cite{1998ApJ...503..325A} with an
optimal convolution kernel modeled as the superposition of a set of
Gaussian kernel base functions which are modulated by a
two-dimensional polynomial function in the x- and y-direction. We use
the standard kernel model with 4 base functions having standard
deviations of 9, 6, 3 and 0.1 pixel and a polynomial order of 2, 4, 6
and 0 respectively. The kernel is normalized (flux conserving) and has
a size of 51$\times$51 pixel. The total number of free parameters is
50. In order to account for PSF variations across the skycell we
divide the image into 8$\times$8 sub-fields and calculate a convolution
kernel for each sub-field separately. Our procedure is 
similar to \cite{2010MNRAS.409..247K}. The same procedure 
is also used for the WeCAPP data-set (Riffeser et al. in prep.).

\begin{figure}[htbp]
  \begin{center}
    \includegraphics[scale=0.3]{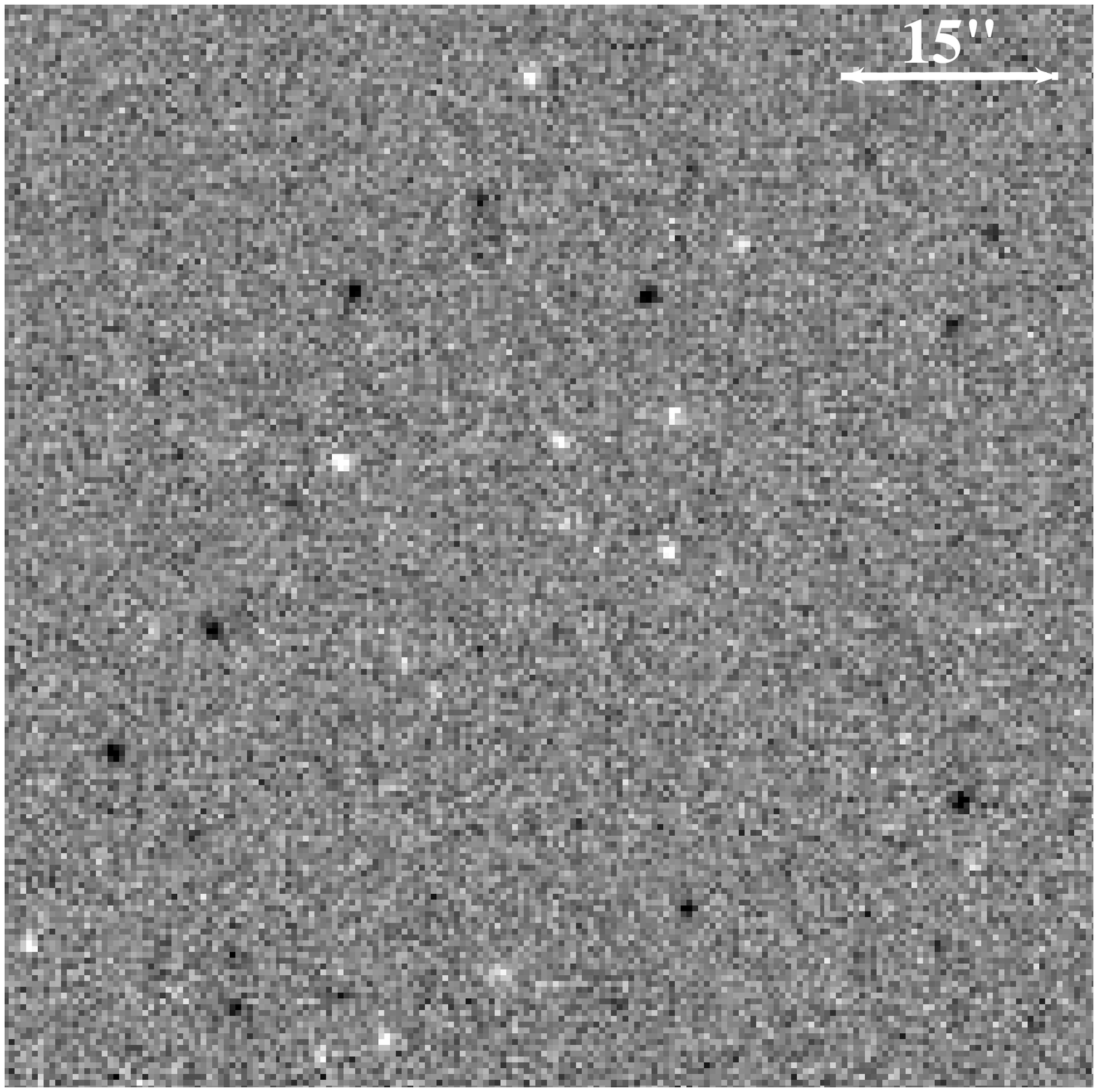}
    \includegraphics[scale=0.3]{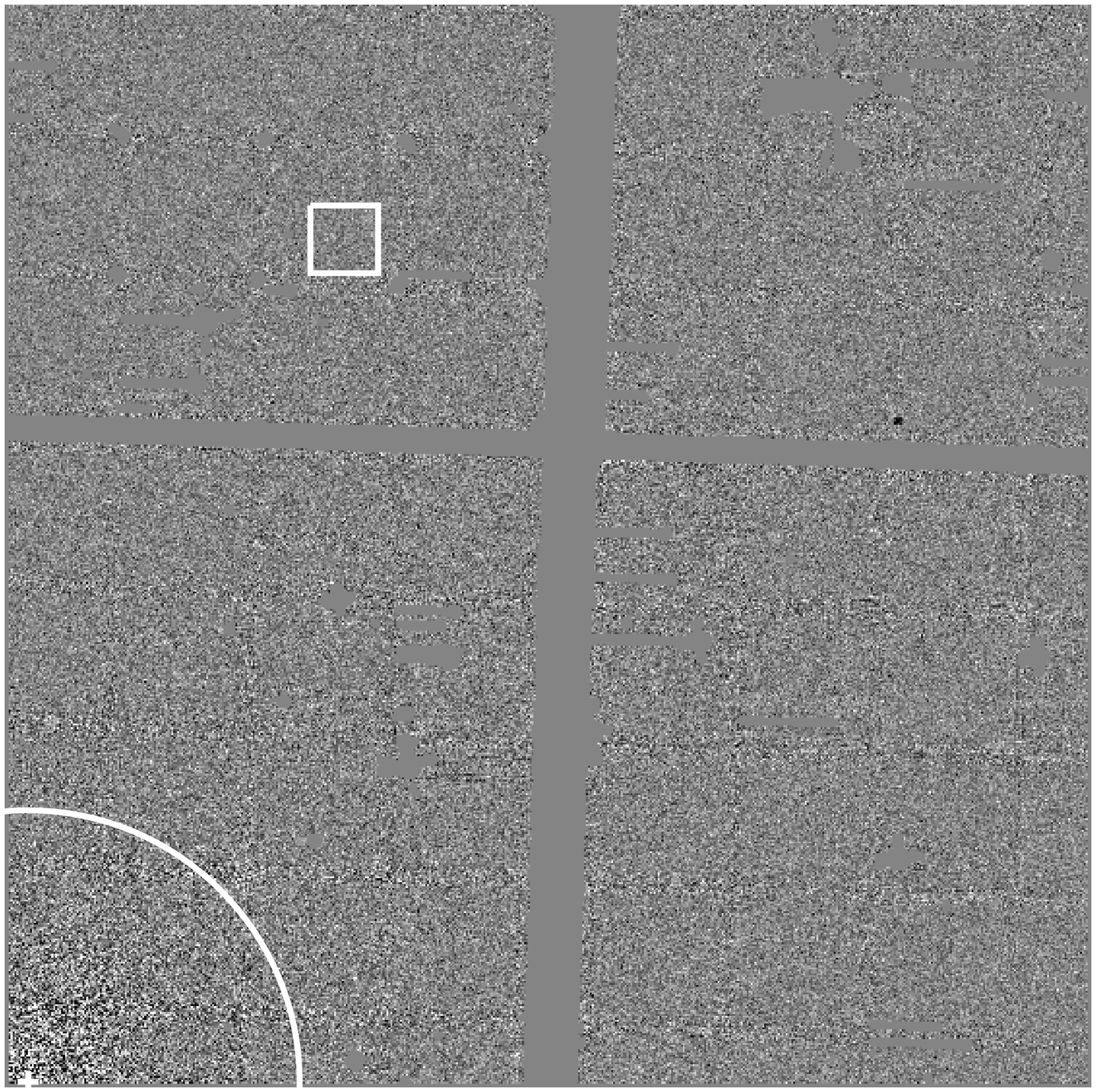}
    \caption{Zoom-ins to an example difference image of skycell 078 in
      $\ips$-band. 
      Left panel: Positive (white) and negative (black) residuals are visible at 
      the position of variable sources.  
      Right panel: The center of M31. The white circle indicates 
      a distance of $5'$ (1.12 kpc) from the center of 
      M31 (marked with the white cross). The vertical and horizontal
      masked area are the gaps between the detectors. The size of this image 
      is $20' \times 20'$. The white box outlines the size of the zoom-in presented in the left panel.
        The flux cut-level of these two images are the same.} 
    \label{fig.diff}
  \end{center}
\end{figure}

The difference images contain only noise at the positions of constant
sources. Variable objects such as microlensing events or eclipsing binaries are
visible as positive or negative PSF-shaped residuals. Fig.
\ref{fig.diff} shows an example difference image of the bulge region
of M31 in which several variable objects are clearly visible. Note
that in the bulge of M31 the stellar density is so high that a large
fraction of the pixels contains variable sources.
\subsection{Creation of the light curves}
\label{sec.lightcurves}

The light curves are created using PSF-photometry on the difference
images. The PSF is constructed using a set of isolated stars in the
convolved reference image. The pixels of each PSF reference star are
subdivided into a sub-pixel grid, shifted and combined. Since the
stars are all on different sub-pixel positions, we obtain a smooth
profile for the combined PSF with sub-pixel precision. The PSF is then
used to measure the flux in each difference image after binning to the
actual sub-pixel position of each star.

As shown in Fig. \ref{fig.flowchart} we produce two sets of light
curves. On the one hand we create a light curve of each pixel which we
use for the detection of microlensing events. These light curves only
contain differential fluxes as measured in the difference images. On
the other hand we create light curves of resolved sources both in M31
and in the foreground. For those light curves we add the flux as
measured in the reference image (see section \ref{sec.refimage}) to
the differential fluxes and obtain absolute fluxes.

\subsection{Flux calibration}
\label{sec.calibration}

The flux calibration requires an existing catalog with similar
filter as PS1.  Thus we use the catalog of Data Release 7 from
SDSS-II \citep{2009ApJS..182..543A}.  The $\rps$ and $\ips$ are rather
close to the SDSS analogs thus the overall flux calibration should
be insensitive to color terms \citep{psphot}.  The data used consist
of the M31 footprints Run 3356, 3366, 3367, 6426 and 7210 with a total
of 6061956 sources.  We calibrate our instrumental magnitude
($m_{instr,\mathrm{PSF}}$) in the reference image to the SDSS
magnitude ($m_{_{SDSS}}$) with the following formula:
\\
\begin{equation}
\begin{array}{rl}
m_{_{SDSS}} & = m_{instr,PSF} + ZP_{\mathrm{PSF}} \\
 &       = -2.5\log(F_{ref}[\mathrm{PSF}_{ref}]) + ZP_{\mathrm{PSF}}
\end{array}
\label{eq.calib_ref}
\end{equation}
where the $F_{ref}[\mathrm{PSF}_{ref}]$ is the flux of a source
determined with the PSF of the reference image PSF$_{ref}$. Note that
the PSF profile and the amplitude is arbitrary and thus different in
each skycell. This is because the reference images are created out of
30 images having the best seeing, which do not necessarily have the
same timestamp. The PSF profile is measured with selected stars on the
reference image.  ZP$_{\mathrm{PSF}}$ takes into account the
difference between the SDSS magnitude and $m_{instr,\mathrm{PSF}}$ and
also depends on the exact PSF-construction (how many and which stars
are used, which kind of data enter the reference image). The flux
calibration is performed on the reference images instead of the
individual image, because the reference images have the highest S/N.

Each difference image is calibrated to have the same flux unit as the
reference image and therefore\\ \footnotesize
\begin{equation} m_{_{SDSS}} = -2.5\log(F_{ref}[\mathrm{PSF}_{ref}] +
\Delta F_{diff}[\mathrm{PSF}_{diff}]) + ZP_{\mathrm{PSF}} \quad .
\label{eq.calib_diff}
\end{equation} \normalsize holds. This is very useful to determine the
magnitude of variables. In an extreme case, the magnitude of a
transient event which has not been detected before
($F_{ref}[\mathrm{PSF}_{ref}]$ = 0) can be easily determined from the
flux in the difference image alone.

The relative calibration of PS1 data with SDSS data in the $\ips$ and
$\rps$ bands does not take into account color terms, but still is
fairly accurate. The reason for this is that these color terms are
small in the $\rps$ and $\ips$ filters \citep[see][]{psphot}. At a
later stage we will nevertheless calibrate our PAndromeda data using
PS1 calibration tools and data directly. In case the reader is 
interested in the magnitude transformation to the Johnson/Kron-Cousins
$BVRI$ system, we refer to the work by \cite{2007ASPC..364..165I}:
\begin{equation}
\begin{array}{l}
R-r_{SDSS} = -0.0107(r_{SDSS}-i_{SDSS})^3 + 0.005(r_{SDSS}-i_{SDSS})^2 -0.2689(r_{SDSS}-i_{SDSS}) - 0.154\\
I-i_{SDSS} = -0.0307(r_{SDSS}-i_{SDSS})^3 + 0.1163(r_{SDSS}-i_{SDSS})^2 - 0.3341(r_{SDSS}-i_{SDSS}) - 0.3584
\end{array}
\label{eq.color}
\end{equation}
with the root-mean-square scatter for residuals evaluated for all their sample stars to be 15 millimag in $R-r_{SDSS}$ and 19 millimag in $I-i_{SDSS}$.

Our single frame data are remapped  twice: there is a remapping to the
sky tessellation where the pixelsize is decreased. Later on when the single
frame prereduced data from IPP (Sect. 3) get ingested into our pipeline the
images are rebinned to twice the pixelsize (Sect. 3.1). The remapping causes
a correlated noise depending also on the interpolation method used. Whereas
our pipeline uses per pixel error propagation for all steps that are done
after ingestion into our pipeline, it does not account for correlated noise
that comes from the resampling before.
We therefore empirically rescale the errors in the lightcurves directly,
such that the formal PSF flux errors match  the scatter of the PSF flux
measurements. This rescaling factor is (as expected) the same for every
skycell and every filter.

\subsection{Microlensing event detection}
\label{sec.detection}
The microlensing event detection is performed on the light curve of
each pixel with successive criteria.  The applied criteria for
detection are as follows. For a preselection, we require at least
three successive flux measurements in the light curve to be 3$\sigma$
above the base-line in either $\rps$ or $\ips$-band.  We then perform
a microlensing fitting with the following formula
\citep[see][]{2006ApJS..163..225R}
\\
\begin{equation}
\Delta_{F}(t)\approx F_\mathrm{eff} \left[\frac{12(t-t_0)^2}{\tfwhm^2}+1\right] + a_0 + a_1(t-t_0)
\label{eq.gould}
\end{equation}
where $F_\mathrm{eff}$ is the effective flux and for high magnifications is similar to the 
flux excess
\begin{equation}
F_\mathrm{eff} := \frac{F_0}{u_0} \approx \Delta_{F} \quad ,
\end{equation}
$t_0$ is the time of flux maximum, 
$\tfwhm$ is the full-width half-maximum event timescale.

The term $a_0$ takes into account the shift of the base-line flux by
a constant in cases for which there is a variable spatially close to the
microlensing event and data of these variable phases enter the
reference image.  The term $a_1$ is to include the gradient in the
base-line flux in case the base-line flux is influenced by a long
period variable close to the microlensing event.  We require the data
point closest to the time of the fitted maximum to have a good PSF.

We then filter out light curves with a $\chi =
\sqrt{\chi^2_\mathrm{dof}}$ (where $\chi^2_\mathrm{dof}$ is the total
$\chi^2$ divided by the degree of freedom) larger than 1.4 in
$\rps$-band.  We also require the S/N of the closest flux measurement
to the time of the fitted maximum to be larger than 9 in both $\rps$
and $\ips$-band.

The light curve of a microlensing event consists of many noisy data
points that scatter around the baseline if one excludes the small
time interval (of order $2 \times \tfwhm$) where the microlensing
event is measurable. The frequency of the deviations from the baseline
depends on their amplitude and they occur at a rate given by the
statistics of the PSF-flux errors. These deviations should be spread
over the survey time and not be clustered in some time intervals.  A
correlated deviation from the baseline can be a hint of a nearby
variable source or a time correlated artefact of data reduction. We
would like to identify such correlations in the light curves and sort
out microlensing candidates that have a good light curve fit but a
signature for a time correlated deviation from the baseline not
connected to the microlensing event.  \hfill\break The
$\chi^2_\mathrm{dof}$ value of a light curve fit only tells how well a
model describes the data averaged over all data points, but it is not
sensitive of ``how" the $\chi ^2$ contributions are spread over the
data points, i.e. the survey time interval. We therefore add another
criterion that quantifies the temporal correlations of the model
mismatch.  We define the model mismatch or deviation vector $ d_i =
f_i - f(t_i) $ where $f_i$ is the $i$-th flux measurement at time
$t_i$ and $f(t_i)$ is the model flux at time $t_i$ according to 
equation (\ref{eq.gould}). The number of data points is $n$.  With the
further definition of $ s_i = \mathrm{sign} \left\{ d_i \right\} $, so
that $s_i=1$ or $s_i=-1$ holds, we introduce
\begin{equation} E_{n} =\sum_{i=1}^{n-1} s_i s_{i+1} \quad .
\label{eq.isingmodel}
\end{equation} $E_{n}$ is a measure of how strongly the deviations of
the data from the model are correlated in time. This equation is known
also as the energy of a Lenz-Ising model \citep{REFERENCE-A} with
nearest neighbor spin interactions only.  $s_i$ is then the spin, and
the spin-spin coupling constant usually called $J_{ij}$ is equal to
minus one in our case.  In neuroscience the Ising model is often used
to describe the energy of an ensemble of neurons \citep{REFERENCE-B}
and this also our motivation for choosing such an extra
term. Equation (\ref{eq.isingmodel}) quantifies (similar from our perspective)
the ``energy" for the deviations of data points from a model as seen by
neurons.

For a random process, i.e. if the deviation vector $d$ has no time
correlation, the expectation value of equation (\ref{eq.isingmodel}) is
zero. The distribution of $E_{n}$ is gaussian with a width of
$\sqrt{n}$.  In order to make this scatter of $E_n$ in
equation (\ref{eq.isingmodel}) independent of the number of data points
(similarly as the $\chi^2_\mathrm{dof}$ is independent of the degrees
of freedom) we normalize equation (\ref{eq.isingmodel}) to
\begin{equation} E= {E_{n}/{\sqrt{ n}}} \quad .
\label{eq.isingmodel_2}
\end{equation} We empirically require $|E|<$ 3 for the full light
curve and $|E^{\rm peak}_n|<$ 3.5 for the 20 flux measurements closest
to the time of the flux maximum. We allow the time correlation of the
deviation vector to be larger around the time of maximum
brightness. This is purely empirical. It can be justified by the need
to allow for time coherent deviations from the
points-mass-point-source standard model in the presence of extended
sources and lenses, and disturbed deflection potentials.
The criteria above (which are summarized below in Table
\ref{tab.MACHOdetection}) are not fine-tuned to pick some particular
PS1 microlensing candidates but they where developed and tested for
the WeCAPP project (Riffeser et al. in preparation).

If the energy criteria $|E|<$ 3 and $|E^{\rm peak}_n|<$ 3.5 are
fulfilled, we select the light curves with $\tfwhm$ shorter than 20
days. This is to avoid the misinterpretation of long period variables
as microlensing events due to the short time span of the first
PAndromeda season. Contamination from the long period variables is
less severe for campaigns with longer time span, e.g. the WeCAPP
project (Riffeser et al. in preparation). All the criteria used are
summarized in Table \ref{tab.MACHOdetection}.  All light curves that
pass the microlensing event detection criteria are examined by eye.
We exclude light curves for which the best-fitted maximum $t_0$ and the
time-interval plus/minus $\tfwhm$ around the maximum are not in the
time-span of the PAndromeda data-set (event time selection).  We also
exclude the light curves with the signature of variability outside
the event timescale.  The latter happens, e.g., when there are
variables or artefacts close to the microlensing candidate that
contribute some flux to the examined PSF light curve.  The number of
event candidates that are discarded after the event time selection is
of the same order as the number of our final events. Thus our
detection method is mostly automatic.
\\
\begin{table}[!h]
\centering
\caption{Microlensing event detection criteria}
\begin{tabular}{cl}
\hline
     & Criterion \\
\hline
I & Three successive 3$\sigma$ flux measurement in $\rps$ or $\ips$  \\
II & Good PSF around $t_0$ \\
III & $\chi_{\mathrm{dof}} <$ 1.4 in $\rps$ \\
IV & S/N $>$ 9 for maximum flux measurement in $\rps$ and $\ips$ \\
V & $|E|<$ 3 for the full light curve \\
VI & $|E^{\rm peak}_n|<$ 3.5 for measurements around $t_0$ \\
VII & $\tfwhm$ $<$ 20 days\\
\hline
\end{tabular}
\label{tab.MACHOdetection}
\end{table}

\subsection{Astrometric precision of the warped images}
\label{sec.astrometry}
In order to achieve high photometric accuracy with the image
differencing technique it is very important to have a good alignment
between the reference image and the single images - in other words to
have a good relative astrometric calibration. In this section we study
the astrometric precision of the PAndromeda warped images and check
the quality of the astrometry achieved by the IPP by calibrating
against 2MASS.

\begin{figure}[htbp]
  \begin{center}
    \includegraphics[scale=0.45]{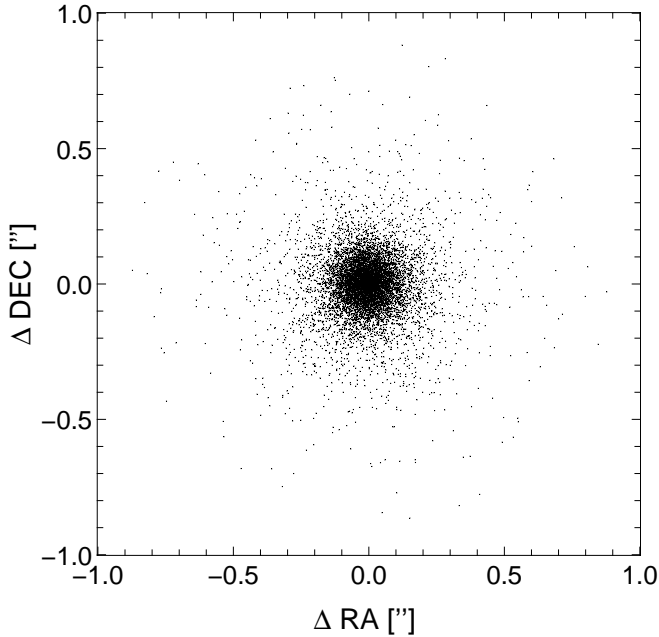}
    \caption{Internal astrometric precision measured by comparing the
      positions of all sources in the $\rps$- and $\ips$-band
      reference images of skycell 068.}
    \label{fig.refref}
  \end{center}
\end{figure}

\begin{figure}[htbp]
  \begin{center}
    \includegraphics[scale=0.45]{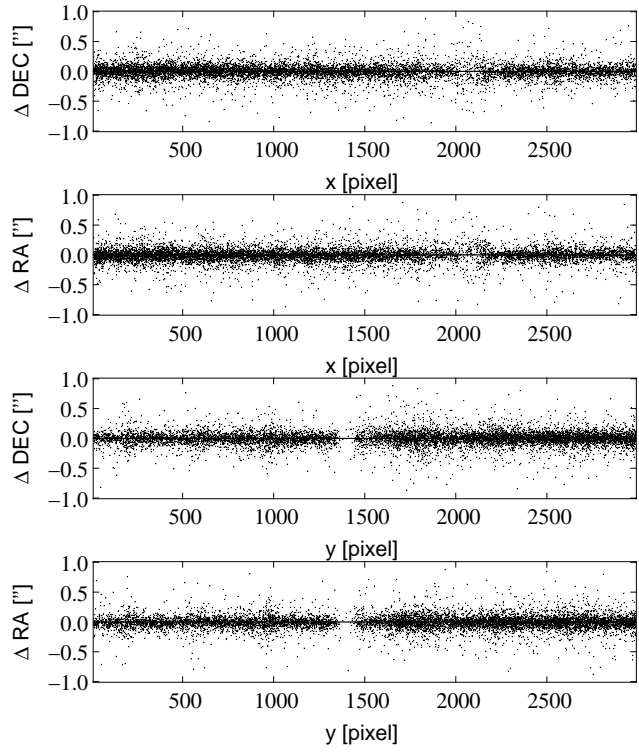}
    \caption{Same astrometric residuals as shown in Fig.
      \ref{fig.refref} plotted as a function of position within the
      skycell. The gaps between the 4 detectors that overlap with this
      skycell are clearly visible (compare Fig. \ref{fig.ref}).}
    \label{fig.refref2}
  \end{center}
\end{figure}

As a first test we compare the x- and y-positions of sources in the
$\rps$- and $\ips$-band reference images of skycell 068 (see
Fig. \ref{fig.refref}).  
The RMS of the residuals is 175 mas (note that the distribution of
residuals shows broad wings, and therefore the RMS value is dominated
by large residuals. The standard deviation of a Gaussian fitted to the
core of the distribution of residuals would be much smaller but this
estimate strongly depends on the fitting radius).  Further we check
for systematic trends of the residuals as a function of position
within the skycell (Fig. \ref{fig.refref2}). There is no trend visible
proving a homogeneous astrometric calibration.

The reference images have been constructed using the images with best
image quality. In order to test if the internal astrometric
calibration is good also for the rest of the data we compare the x-
and y-positions of sources in two individual $\rps$-band visit stacks,
one of them having a very low airmass (1.07) and the other one having
a high airmass (1.54). We find a somewhat lower precision (RMS of 237
mas) but no trend with position within the skycell.

\begin{figure}[htbp]
  \begin{center}
    \includegraphics[scale=0.45]{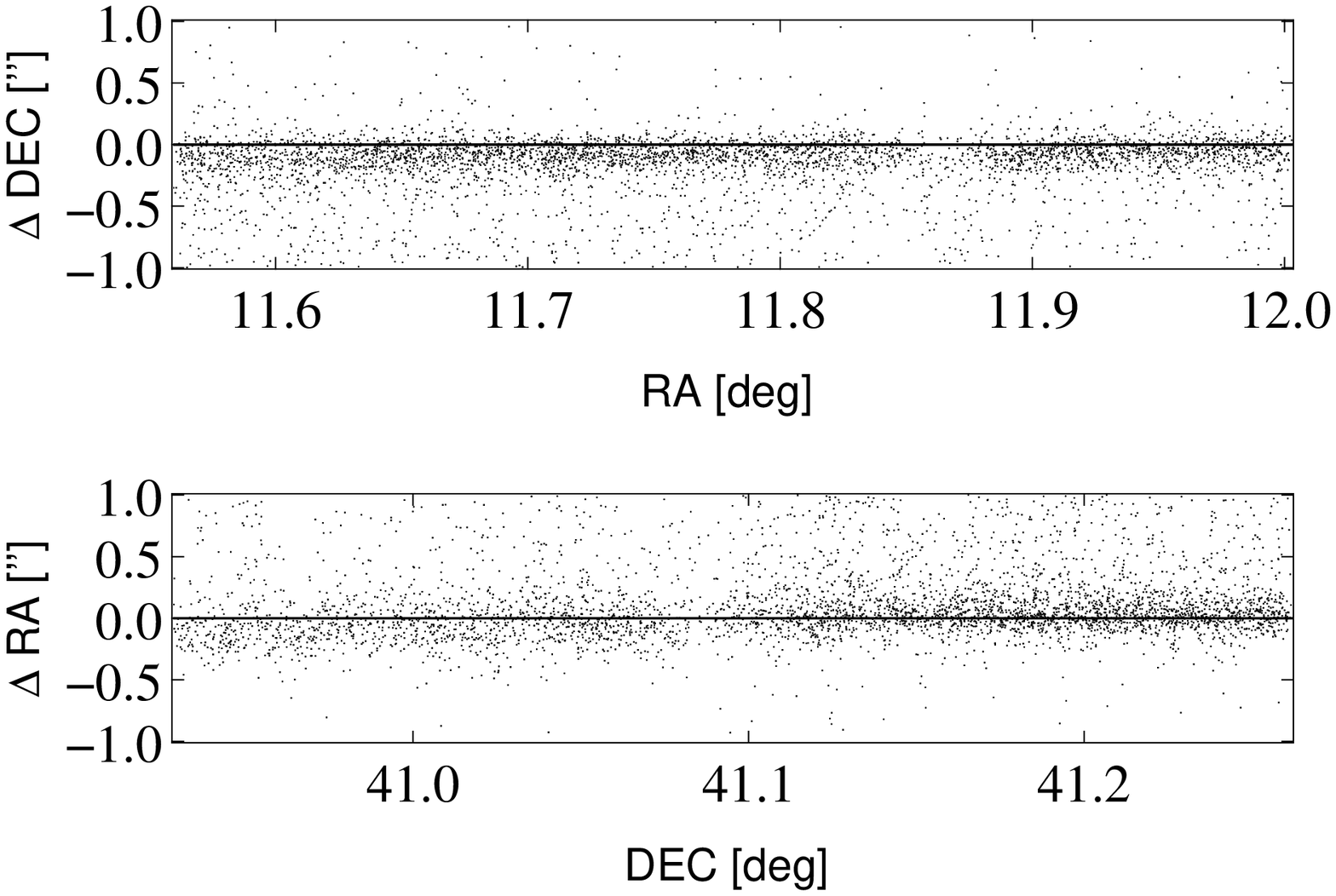}
    \caption{Absolute astrometric precision measured by comparing the
      positions of all sources in the $\rps$-band reference images
      of skycell 068 to a catalog of the Sloan Digital Sky Survey
      DR7. There is a small offset in Declination and a trend in
      Right Ascension as a function of Declination. Both effects are
      at the sub-pixel level ($\sim$ $0.1''$).}
    \label{fig.refsloan}
  \end{center}
\end{figure}

We also tested the absolute astrometric precision comparing the
positions measured in the $\rps$-band reference images of skycell 068
with a catalog from SDSS DR7 \citep{2009ApJS..182..543A}.  Here we see
a systematic trend only at the sub-pixel level in the RA coordinate as
a function of RA as well as a constant systematic offset of the DEC
coordinates (see Fig. \ref{fig.refsloan}).  We cannot say whether this
points to a problem in the astrometric calibration done by the IPP or
if the SDSS catalog has a calibration issue. In any case, for the
quality of the light curves only the relative astrometric precision is
important and since the offset is small compared to the pixel size we
see no problems with identifying objects when cross-matching with
external catalogs.

\subsection{Photometric stability and quality of the light curves}
\label{sec.photometry}

\begin{figure}[!h]
  \centering
  \includegraphics[scale=0.6]{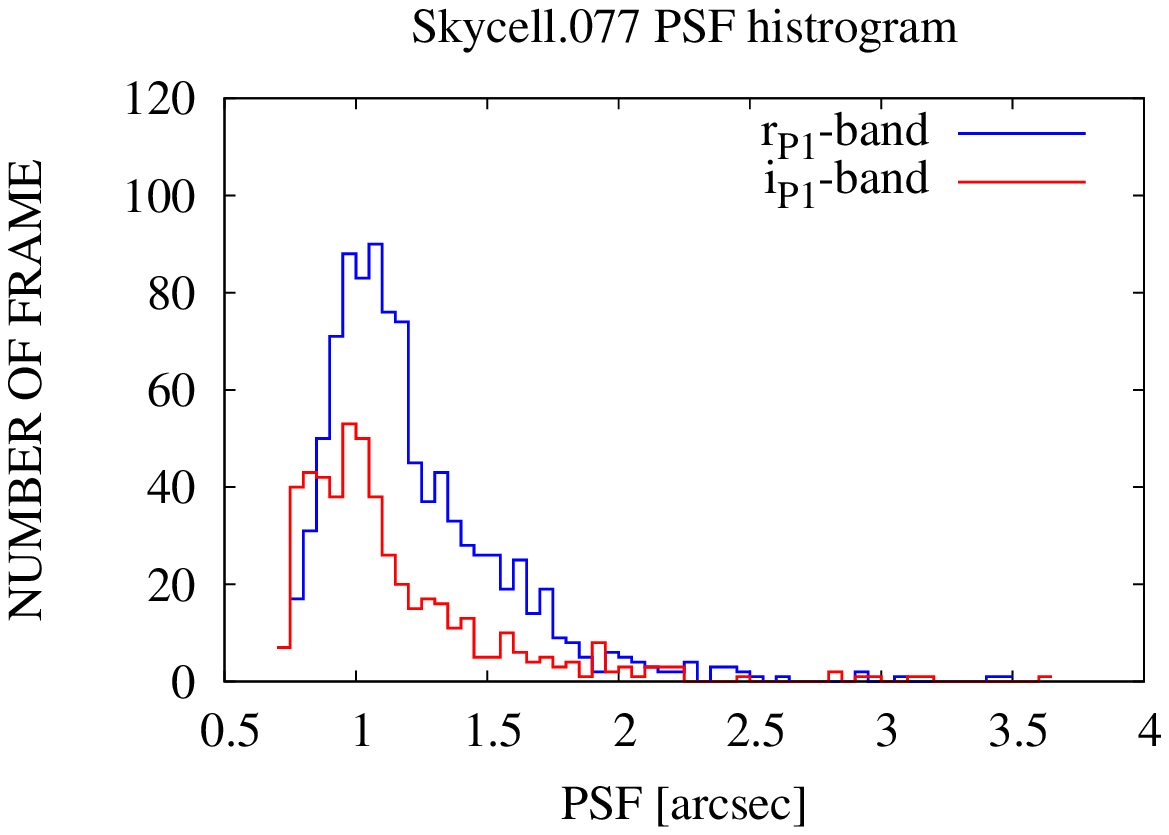}
  \includegraphics[scale=0.6]{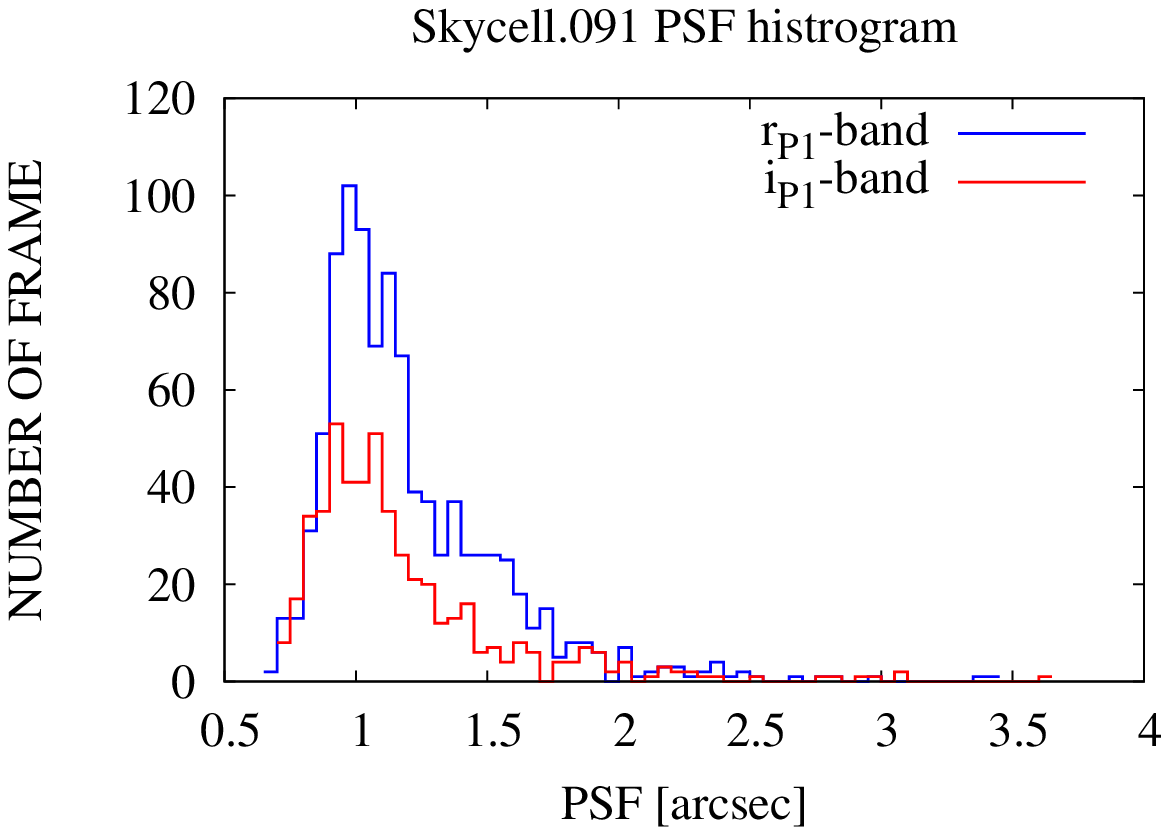}
  \caption{Point-Spread Function (PSF) distribution of the first
season of PAndromeda in the bulge (left panel) and disk (right panel)
field. The PSF distributions of the $\rps$ ($\ips$) data are shown in
the blue (red).}
  \label{fig.psf}
\end{figure}

To understand the data quality of PAndromeda, we first investigate two
sub-fields, one in the bulge of M31 (skycell 077) and one in the disk
(skycell 091). A histogram of the PSF distribution is shown in
Fig. \ref{fig.psf}. The maximum peaks at $1''$.

\begin{figure}[!h]
  \centering
  \includegraphics[scale=0.4]{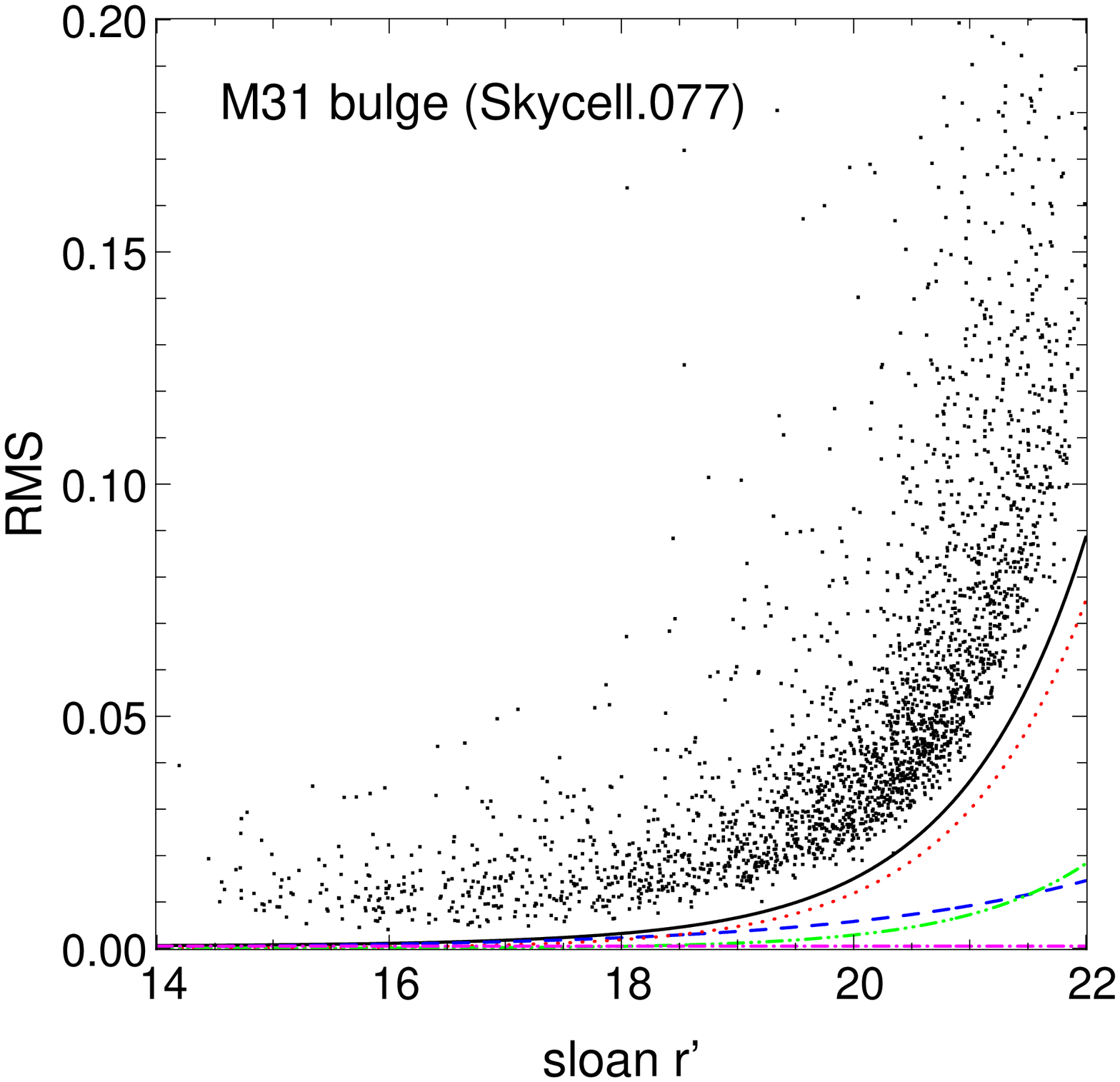}
  \includegraphics[scale=0.4]{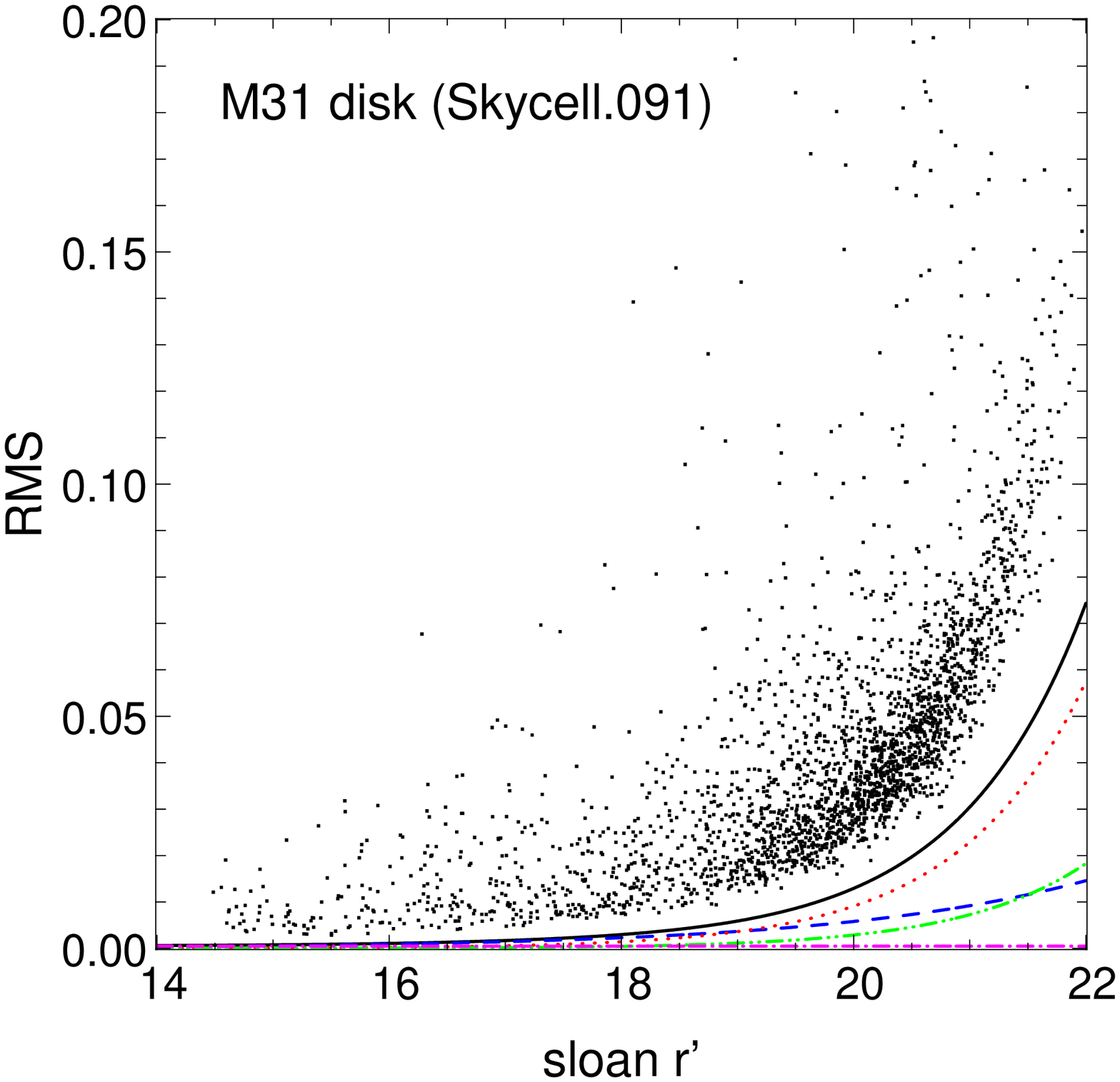}
  \caption{Percentage photometric accuracy (rms) as a function of the
magnitude of stars from images taken by PAndromeda in the bulge field
(left panel) and the disk field (right panel). The RMS is derived from
the light curves of resolved stars. The black line is the expected
total RMS, taking into account the contributions from sky (red line),
read out (green line), object (blue line) and scintillation noise
(magenta line).}
  \label{fig.phot_lc}
\end{figure}

To quantify the precision of the photometry, we perform a test on
sub-fields in the bulge, disk and the astrometric fields.  We use
SExtractor to extract the position of the resolved stars in the
reference images of the sub-fields and create the light curves,
i.e. flux difference of these stars relative to the mean flux as a
function of time, at the position of these stars with our PAndromeda
pipeline. Since most of the stars are not variable, the scatter of the
estimated flux difference (relative to the mean flux) is an estimate
for the photometric precision as a function of the magnitude of the
star. The ratio for this RMS-error relative to the stellar flux is
plotted for each star in Fig. \ref{fig.phot_lc}.  In the plot we
present the 3$\sigma$ clipped RMS value.  The theoretical S/N
predictions, account for the contributions from sky, read out, object
and scintillation noise, are shown in the black lines. We assume the
surface brightness of M31 in $\rps$ to be 20.5 and 21.5 mag/arcsec$^2$ at the
location of skycell 077 and skycell 091
\citep{2011A&A...526A.155T}. The RMS distributions from the data are
shifted from the theoretical expectations. This is because masking is
not taken into account.  Besides, the resampling to the skycells and
binning process also has influenced the quality of the light
curves. Nevertheless, the result shows that the photometric precision
is 1\% for stars with $\rps$ = 19 mag.

\section{Examples for variable light curves}
\label{sec.variable}
\begin{figure}[!h]
  \centering
  \includegraphics[scale=0.8]{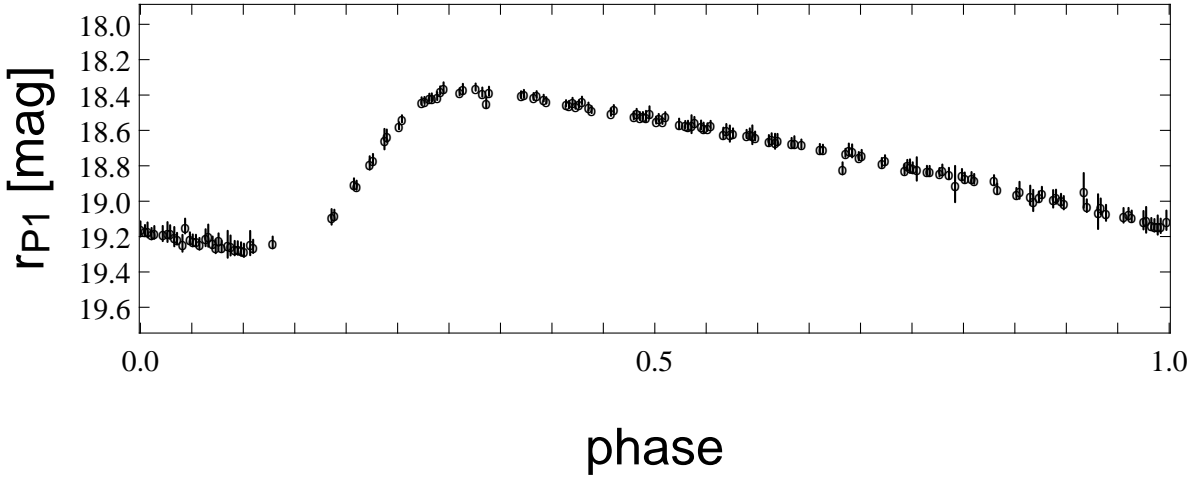}
  \caption{Folded $\rps$-band light curve of our best Cepheid candidate at
    RA(J2000) = 10.2662 deg and Dec(J2000) = 41.1512 deg. The period is 45.6
    days and the distance from the center of M31 is $20'$.}
  \label{fig.DC_lc}
\end{figure}

\begin{figure}[!h]
  \centering
  \includegraphics[scale=0.8]{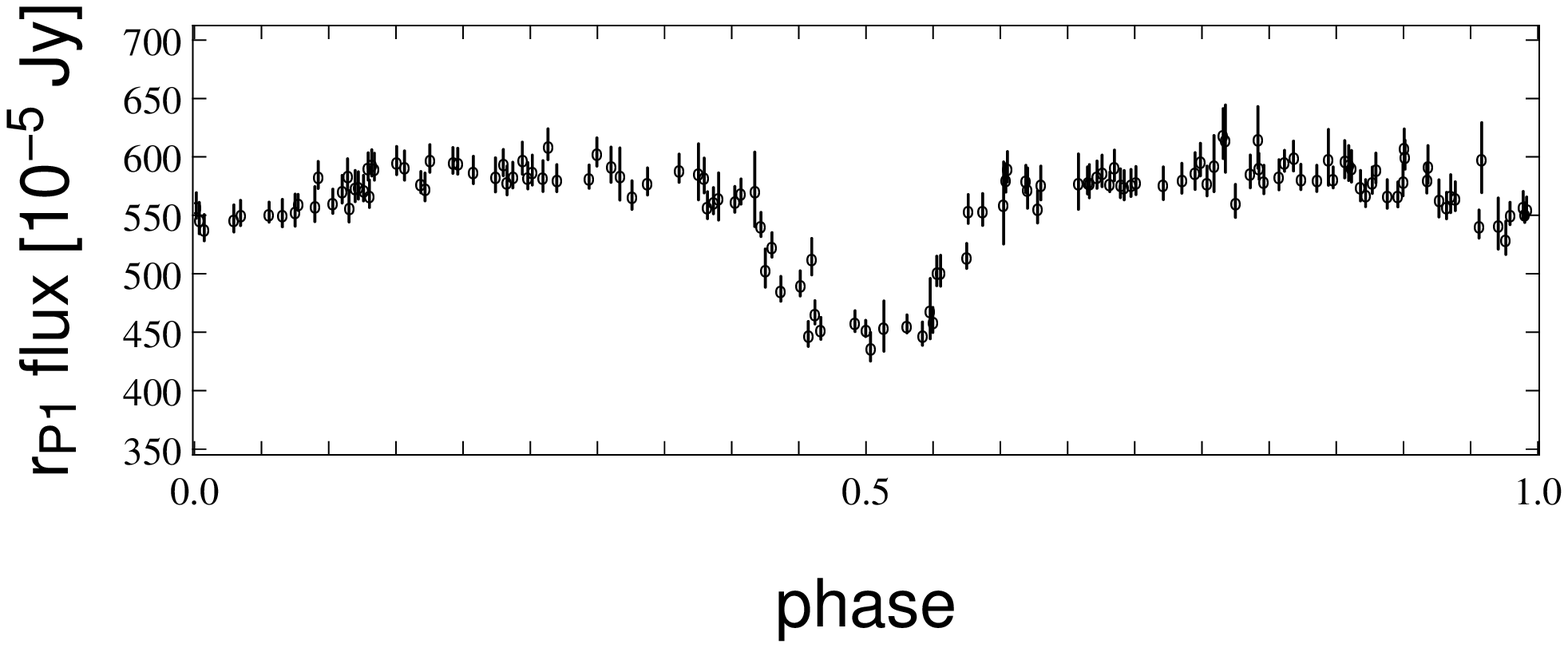}
  \caption{Folded $\rps$-band light curve of a foreground eclipsing binary at RA(J2000) = 10.6698 deg and Dec(J2000) = 41.1681 deg. The period is 0.5 day.}
  \label{fig.EB_lc}
\end{figure}

We also analyze the light curves of the resolved sources in the
central $40' \times 40'$ sub-field. We find about 60 Cepheid-like
sources by inspecting the light curves with large $\chi^2$ when
fitting a constant base-line.  An example light curve is shown in
Fig. \ref{fig.DC_lc}.  These Cepheid-like sources and further
candidates that will be found in the remaining M31 fields will be
analyzed in a forthcoming publication (G\"ossl et al. in prep.).  In
addition, we find one eclipsing binary and present its folded light
curve in Fig. \ref{fig.EB_lc}.

\begin{figure}[!h]
  \centering
\includegraphics[scale=0.8]{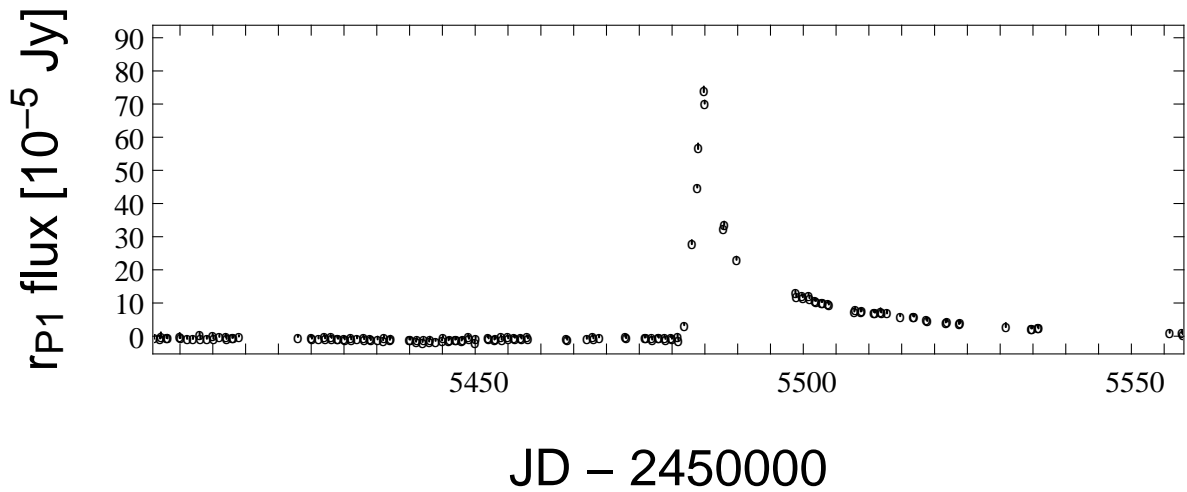}
\caption{$\rps$-band light curve of the nova M31N-2010-10c from PAndromeda survey at RA(J2000) = 11.1108 deg and Dec(J2000) = 41.5205 deg.}
  \label{fig.nova_lc}
\end{figure}

During the time span of the first observing season of PAndromeda we 
find a few nova candidates in the central $40' \times 40'$ sub-field. 
An example light curve is shown in Fig. \ref{fig.nova_lc}. 

\clearpage
\section{The first microlensing events}
\label{sec.ML}

\begin{figure}[!h]
  \centering
  \includegraphics[scale=0.7]{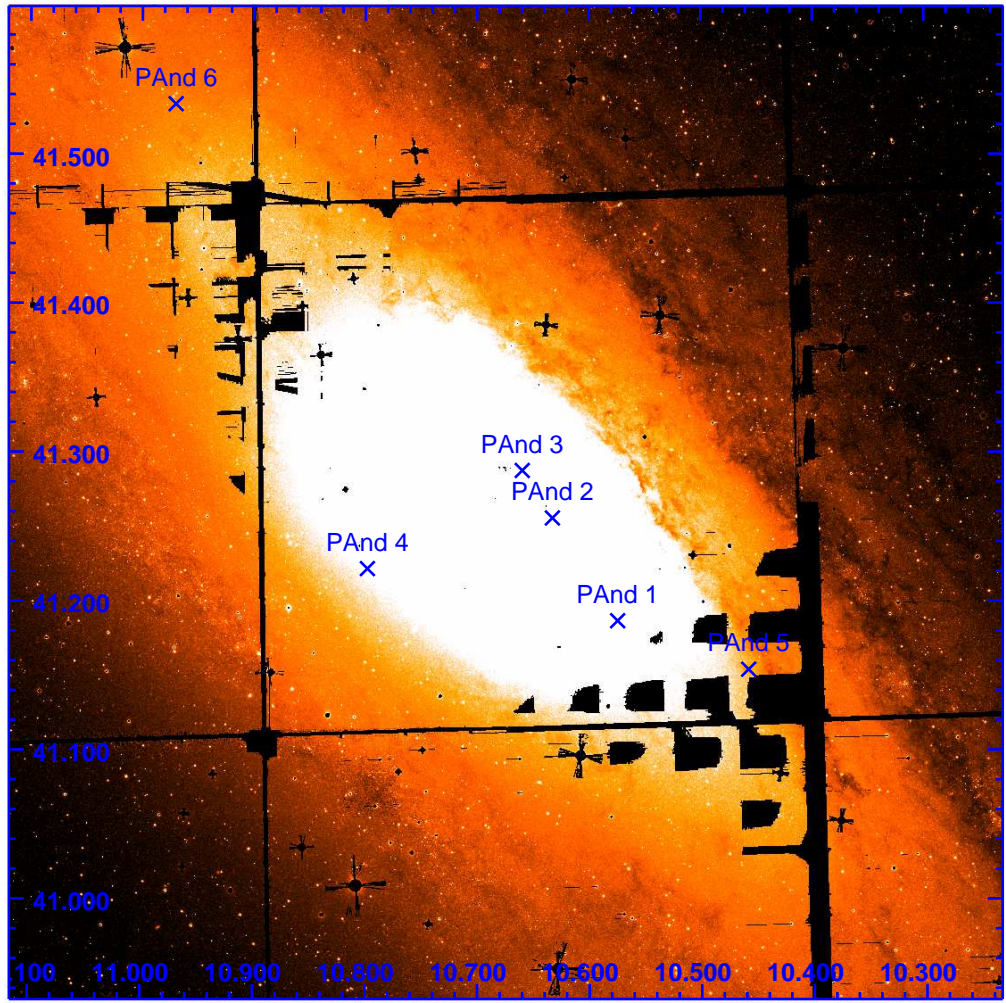}
  \caption{Position of the six microlensing event candidates detected
in the central $40' \times 40'$ region of M31 from PAndromeda. The
coordinates, RA (J2000) in hour and Dec (J2000) in degree are also
shown in the figure. The FOV of this image is $40' \times 40'$.}
  \label{fig.GL_pos}
\end{figure}

\begin{figure}[!h]
  \centering
  \includegraphics[scale=0.4]{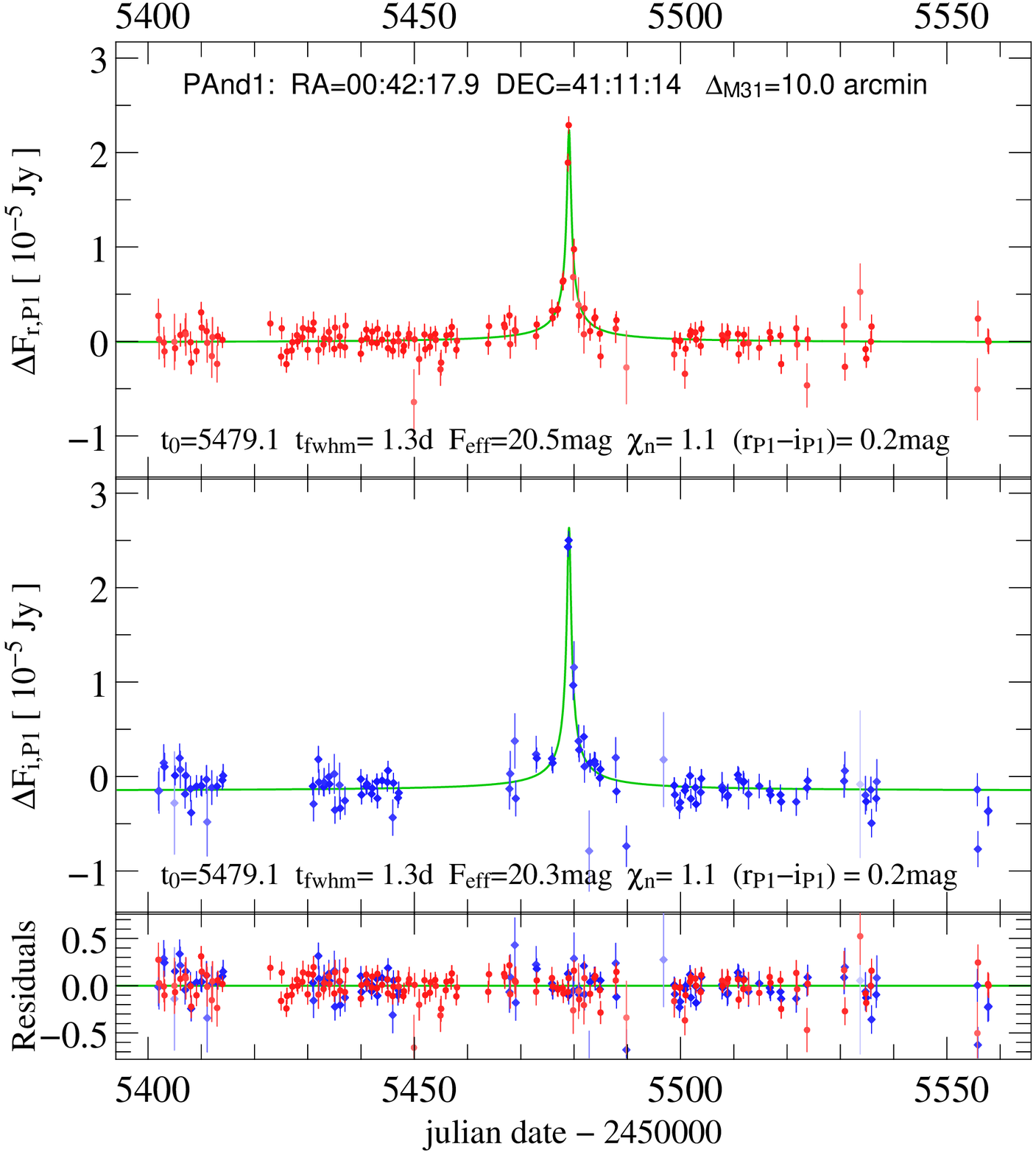}
  \includegraphics[scale=0.4]{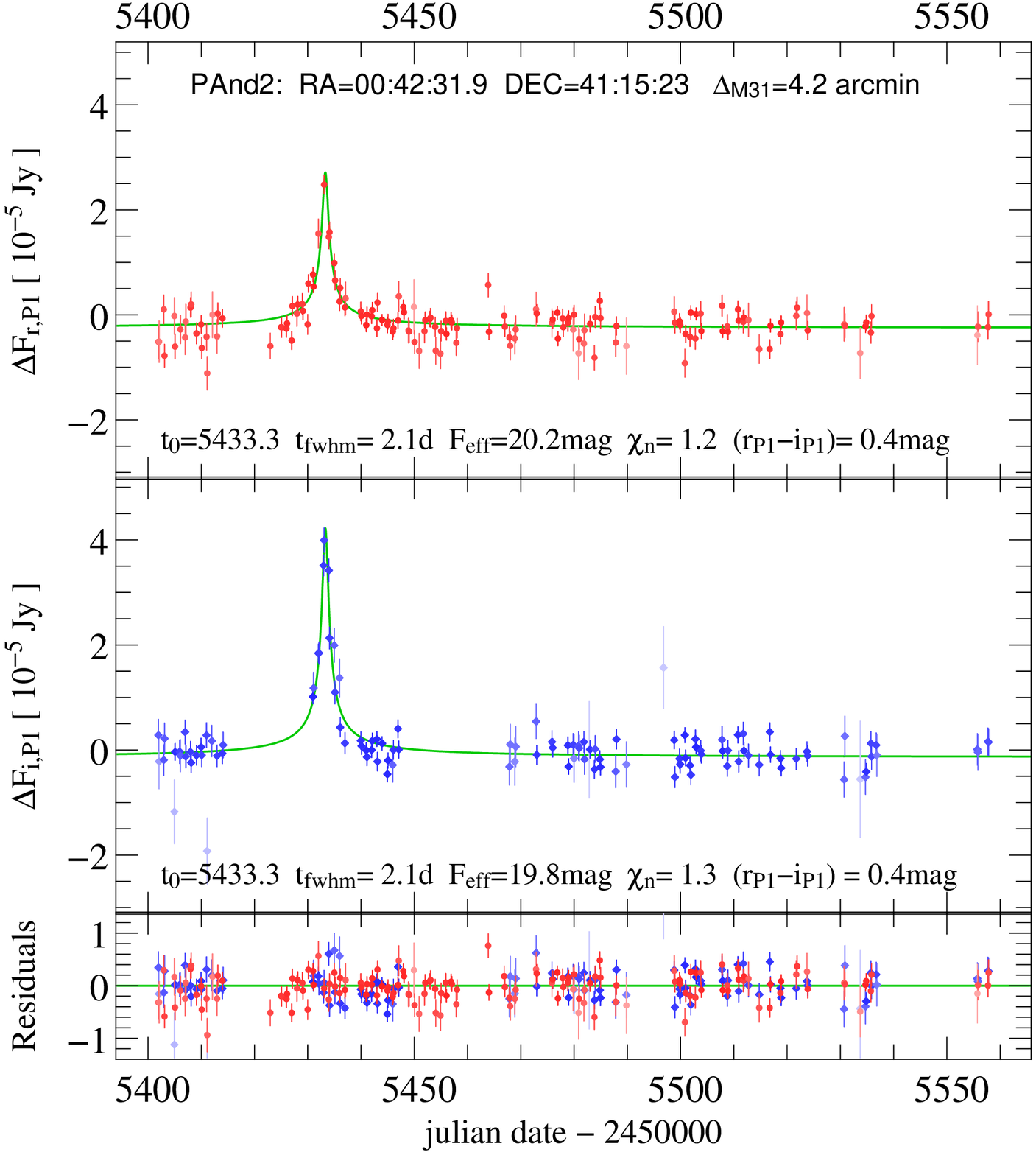}
  \includegraphics[scale=0.4]{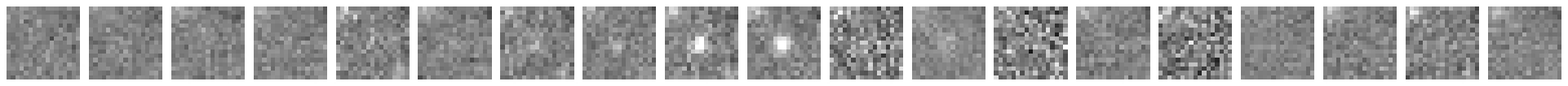}
  \includegraphics[scale=0.4]{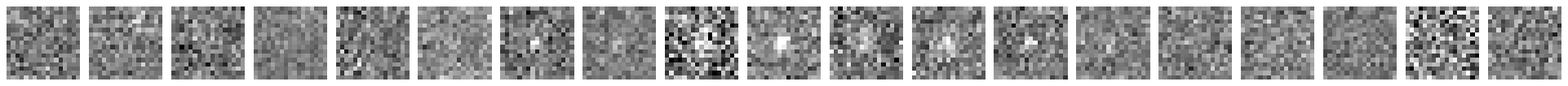}
  \includegraphics[scale=0.4]{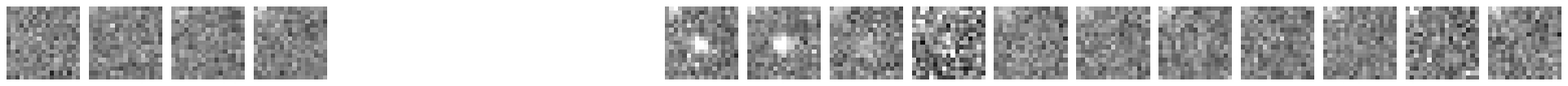}
  \includegraphics[scale=0.4]{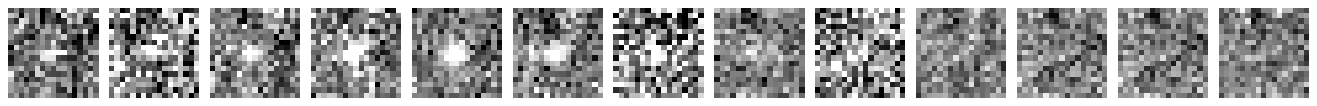}
  \includegraphics[scale=0.4]{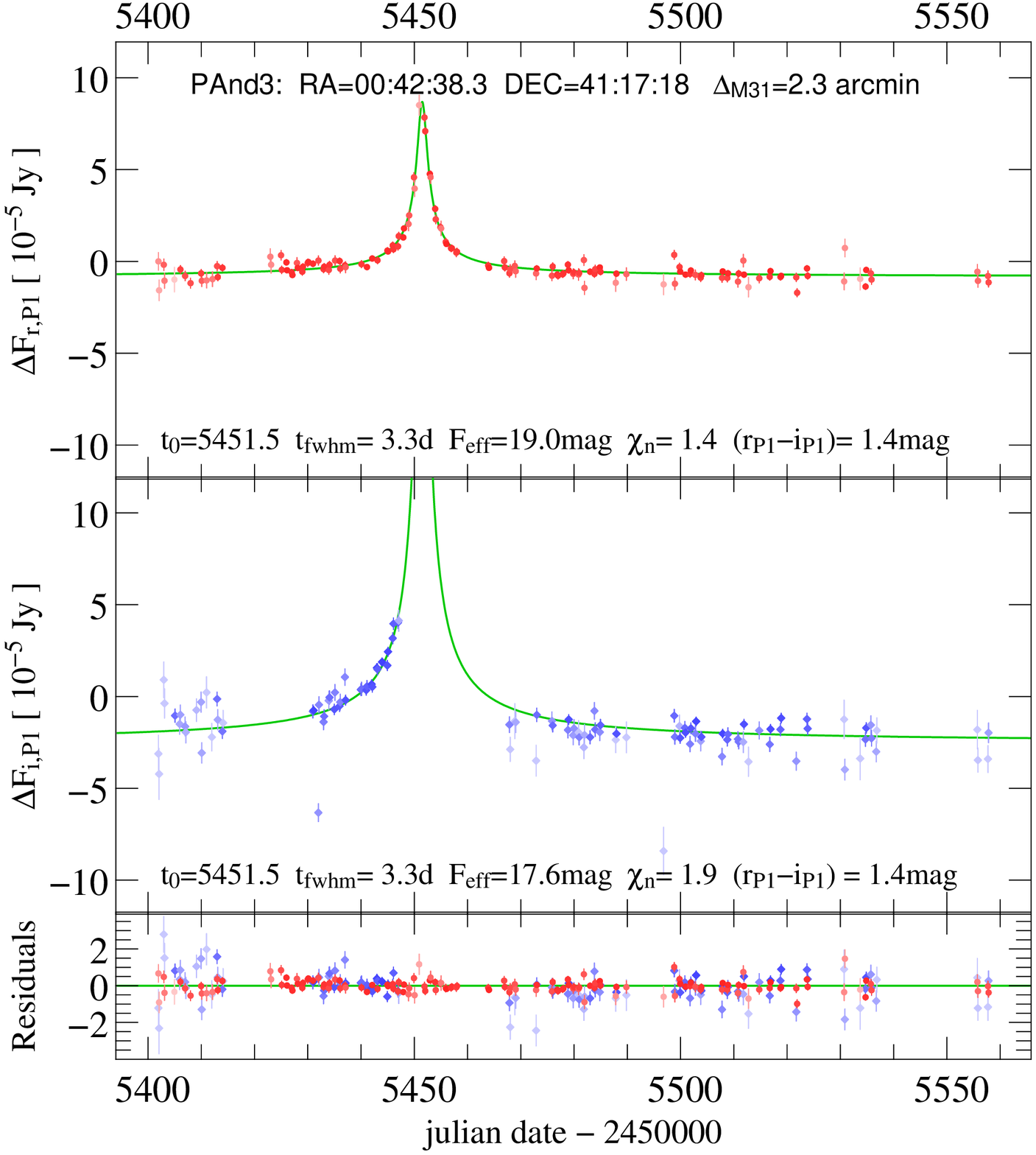}
  \includegraphics[scale=0.4]{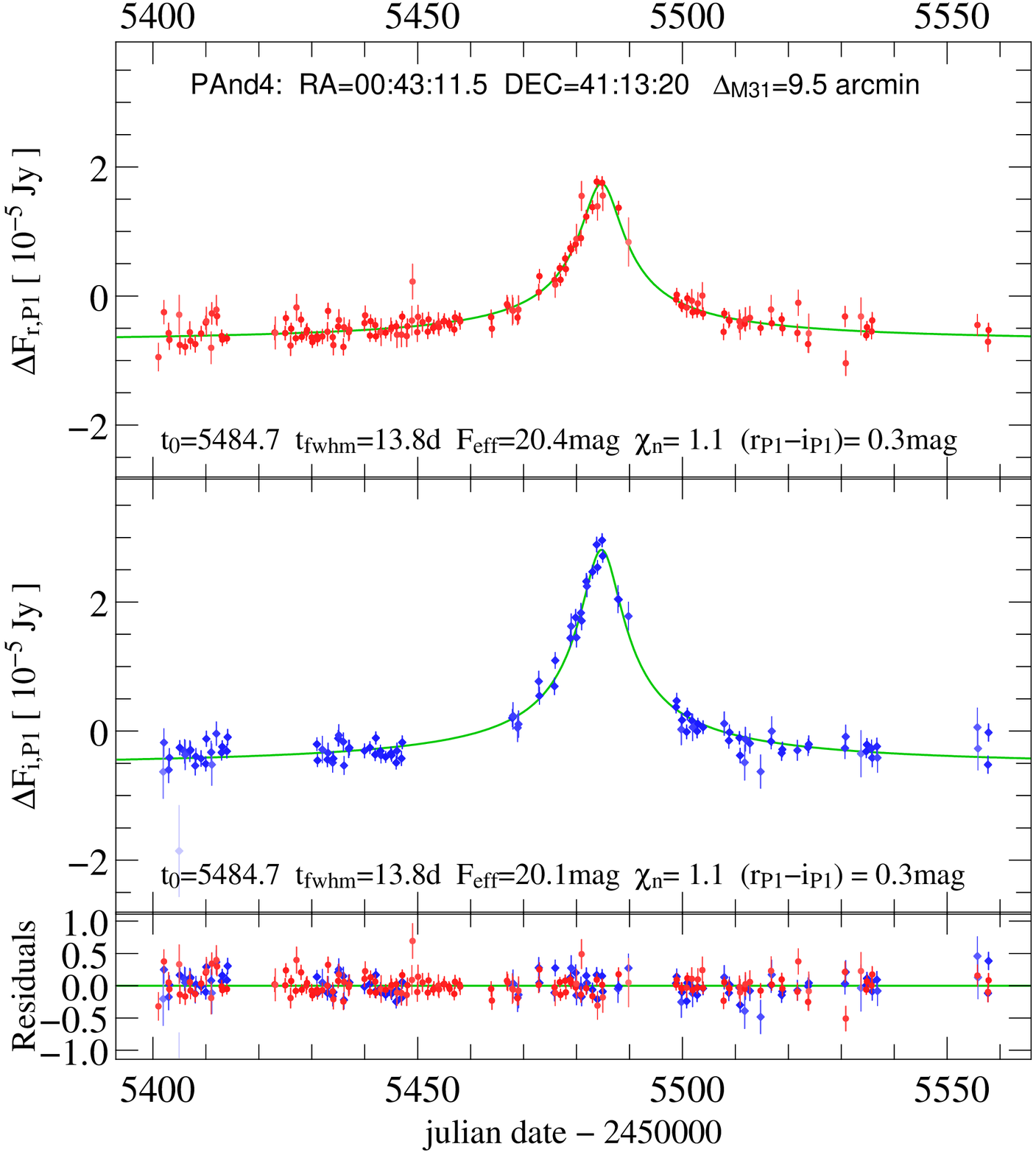}
  \includegraphics[scale=0.4]{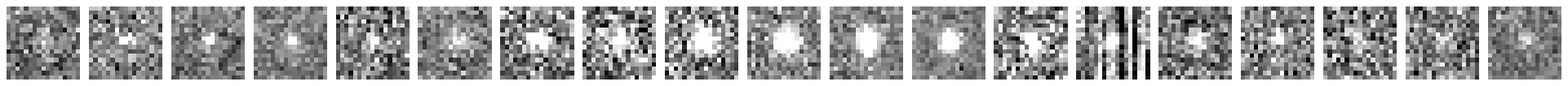}
  \includegraphics[scale=0.4]{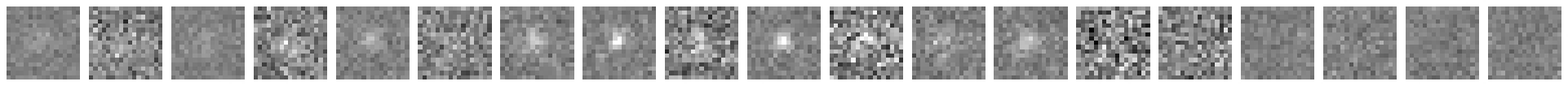}
  \includegraphics[scale=0.4]{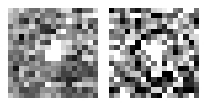}
  \includegraphics[scale=0.4]{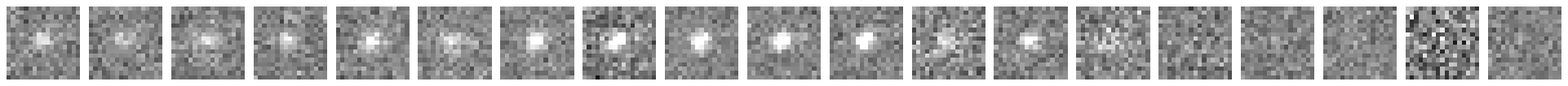}
  \caption{Light curves of the microlensing events detected in the
central $40' \times 40'$ region of M31 from PAndromeda.  Upper (middle)
panel shows the light curve in $\rps$ ($\ips$) and the lower panel 
shows the residuals to the best-fitted light curve. In the figure there 
are also the name of the event (e.g. PAnd-1), the coordinates $\alpha$
(RA) and $\delta$ (Dec) at the epoch of J2000, the distance to the
center of M31 ($\Delta_{\mathrm{M31}}$) in arcminutes. The best-fit
light curves (green) and parameters (black) are also shown, which are the time at
maximum magnification ($t_0$), the event timescale ($\tfwhm$) in units
of a day, the equivalent magnitude at maximum magnification
($F_{\mathrm{eff}}$) in each filter, the normalized $\chi_n$
:=$\sqrt{\chi_{\mathrm{dof}}^2}$ in each filter and the color
($\rps$-$\ips$) of the event.  We also show the postage stamps of the
events ($\rps$ in the upper row and $\ips$ in the lower row). They are
ordered by their observation date, with the flux maximum located in
the center of the sequence.}
  \label{fig.GL_lc1}
\end{figure}

\begin{figure}[!h]
  \centering
  \includegraphics[scale=0.4]{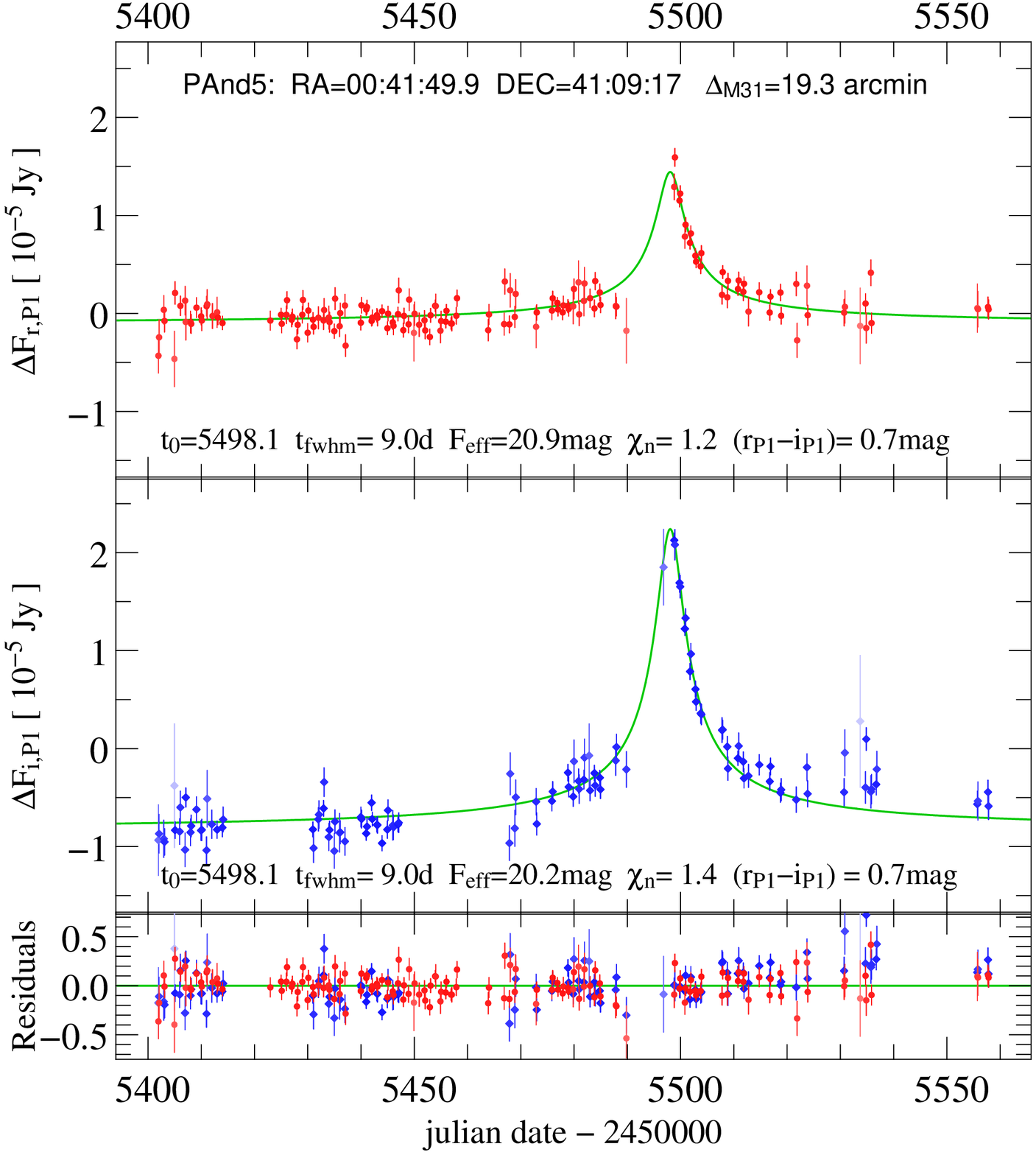}
  \includegraphics[scale=0.4]{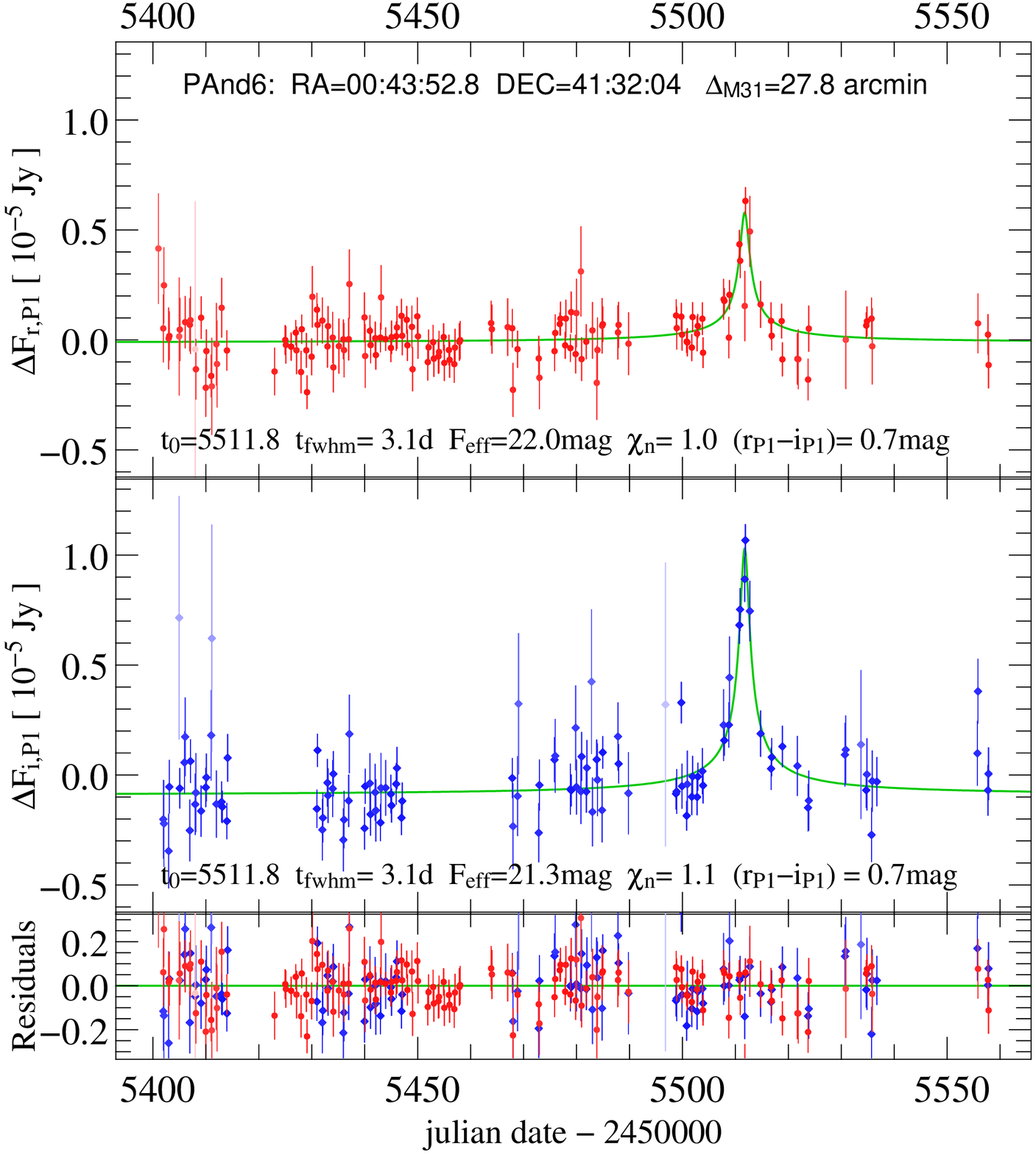}
  \includegraphics[scale=0.4]{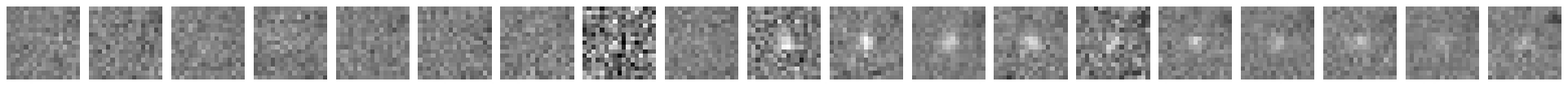}
  \includegraphics[scale=0.4]{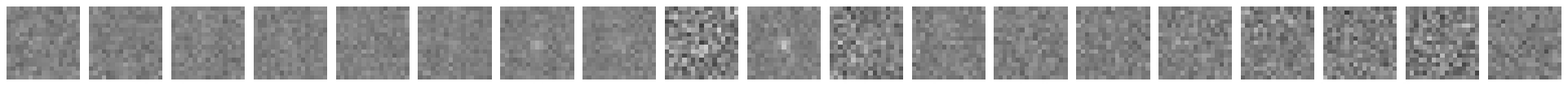}
  \includegraphics[scale=0.4]{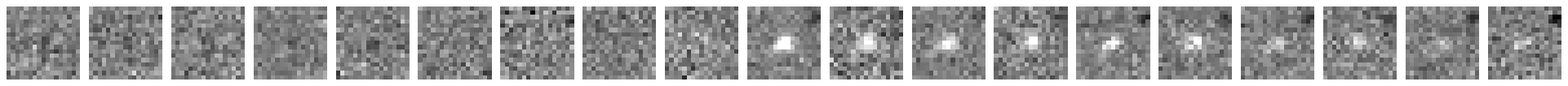}
  \includegraphics[scale=0.4]{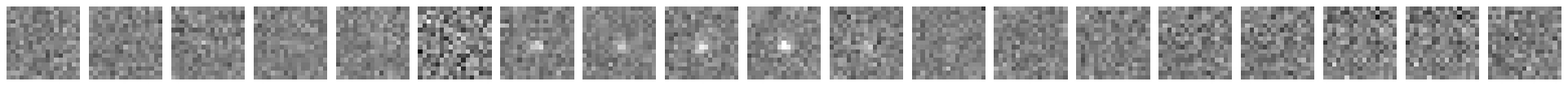}
  \caption{Light curves of PAnd-5 and PAnd-6. 
   See the figure caption of Fig. \ref{fig.GL_lc1} for further explanation.} 
  \label{fig.GL_lc5}
\end{figure}

To compare with the previous results, we first analyzed the central region of the M31 field ($40' \times 40'$) 
and detect 6 candidates. The 
positions and light curves of these 6 candidates are shown in Fig. \ref{fig.GL_pos}, Fig. \ref{fig.GL_lc1} 
and Fig. \ref{fig.GL_lc5}. 
The fitting parameters (as shown in equation (\ref{eq.gould})) and the $\chi_{\mathrm{dof}}$ are presented in Table \ref{tab.ML}.

\clearpage
\footnotesize
\begin{table}[!ht]
\centering
\begin{sideways}
\label{tab.ML}
\begin{minipage}{240mm}
\caption{PAndromeda microlensing event candidates in $\rps$ - and $\ips$-band.}
\begin{tabular}{llllrlllllll}
\hline
    & IAU name & RA(2000)   & DEC(2000) & $\Delta_{\mathrm{M31}}$ & $t_0$         & $\tfwhm$ & $F _\mathrm{eff,\rps}$ & $F _\mathrm{eff,\ips}$  & $(\rps-\ips)$\footnote{The $\rps$ and $\ips$ are rather
similar to the SDSS analogs \citep[see][for further description.]{psphot}.} & $\chi _\mathrm{dof,\rps}$ & $\chi _\mathrm{dof,\ips}$ \\
        & & [h:m:s]    & [d:m:s]   &  &  & [day]          & [ABmag]       & [ABmag]        & [ABmag] &                  &                  \\
\hline
 PAnd-1 & PSO J010.5746+41.1872 & 00:42:17.9 & 41:11:14  & $10.0'$ & 5479.05$\pm$0.04 &  1.28$\pm$0.09 & 20.52$\pm$0.05 & 20.29$\pm$0.06 & 0.23$\pm$0.10  & 1.1 & 1.1 \\
 PAnd-2 & PSO J010.6329+41.2564 & 00:42:31.9 & 41:15:23  &  $4.2'$ & 5433.33$\pm$0.04 &  2.07$\pm$0.16 & 20.22$\pm$0.06 & 19.80$\pm$0.08 & 0.42$\pm$0.13  & 1.2 & 1.3 \\
 PAnd-3 & PSO J010.6596+41.2883 & 00:42:38.3 & 41:17:18  &  $2.3'$ & 5451.49$\pm$0.04 &  3.34$\pm$0.14 & 18.95$\pm$0.03 & 17.60$\pm$0.05 & 1.35$\pm$0.08  & 1.4 & 1.9 \\
 PAnd-4 & PSO J010.7979+41.2222 & 00:43:11.5 & 41:13:20  &  $9.5'$ & 5484.73$\pm$0.11 & 13.82$\pm$0.43 & 20.41$\pm$0.02 & 20.07$\pm$0.03 & 0.34$\pm$0.06  & 1.1 & 1.1 \\
 PAnd-5 & PSO J010.4579+41.1547 & 00:41:49.9 & 41:09:17  & $19.3'$ & 5498.05$\pm$0.17 &  9.04$\pm$0.36 & 20.92$\pm$0.04 & 20.18$\pm$0.05 & 0.74$\pm$0.07  & 1.2 & 1.4 \\
 PAnd-6 & PSO J010.9700+41.5344 & 00:43:52.8 & 41:32:04  & $27.8'$ & 5511.76$\pm$0.12 &  3.08$\pm$0.40 & 21.97$\pm$0.08 & 21.27$\pm$0.12 & 0.69$\pm$0.22  & 1.0 & 1.1 \\
\hline
\multicolumn{12}{l}{We give the IAU name of the 6 microlensing candidates in column 2, the coordinates in columns 3 and 4, the distance to the center of M31}\\ 
\multicolumn{12}{l}{($\Delta_{\mathrm{M31}}$) in column 5, the time of flux maximum (JD - 2450000) and event timescale in columns 6 and 7 the maximum flux in units of}\\
\multicolumn{12}{l}{magnitude in $\rps$ and $\ips$-band in columns 8 and 9, the color in column 10 and the best-fitted $\chi_\mathrm{dof}$ in columns 11 and 12.}\\
\end{tabular}
\end{minipage}
\end{sideways}
\end{table}
\normalsize

\begin{table}[!ht]
\centering
\begin{sideways}
\label{tab.MLRI}
\begin{minipage}{200mm}
\caption{PAndromeda microlensing event candidates with magnitudes in $R$ and $I$-band.}
\begin{tabular}{cccc}
\hline
    & $F _\mathrm{eff,R}$ & $F _\mathrm{eff,I}$  & $(R-I)$ \\
    & [mag]       & [mag]        & [mag]                \\
\hline
PAnd-1 & 20.10 & 19.86 & 0.24\\
PAnd-2 & 19.75 & 19.32 & 0.43\\
PAnd-3 & 18.21 & 16.93 & 1.28\\
PAnd-4 & 19.96 & 19.61 & 0.35\\
PAnd-5 & 20.36 & 19.63 & 0.73\\
PAnd-6 & 21.42 & 20.72 & 0.70\\
\hline
\multicolumn{4}{l}{For readers who are used to the Johnson-Cousins system, we}\\ 
\multicolumn{4}{l}{provide the magnutude and color in $R$ and $I$ from equation (\ref{eq.color}).}\\ 
\end{tabular}
\end{minipage}
\end{sideways}
\end{table}
\normalsize

\clearpage

\clearpage

The 6 PAndromeda events form 3 groups. PAnd-2 and PAnd-3 are at $4.2'$
and $2.3'$ (0.5-1.0 kpc) from the Andromeda center.  This region will be
bulge-bulge lensing dominated regardless of the halo-MACHO
content. The colors ($\rps-\ips$ = 0.4 and 1.4 mag, $R-I$ = 0.4 and 1.3) and short time
scales ($\tfwhm$ = 2.1 and 3.3 day) of the events are in agreement
with bulge stars lensing post main sequence bulge stars
\citep[see][]{2006ApJS..163..225R}.  The events PAnd-4 and PAnd-1 are
at $9.5'$ and $10'$ distance from the M31 center along the minor
and major axes (corresponding to $\sim$ 2 kpc). These are regions
where the light profile turns from bulge light dominated to disk light
dominated \citep[compare, e.g. the Fig. A.1 of][for the $g$-band dust
corrected minor and major M31 surface brightness
profiles]{2011A&A...526A.155T}. Hence, the bulge-bulge and to a
smaller degree the bulge-disk self-lensing rates are reduced relative
to the center. On average the sources of self-lensing become more
likely disk stars, and thus bluer.  The timescales of self-lensing
stay short, i.e. a few days. PAnd-1 has a timescale of 1.3 days, the
other (PAnd-4) is considerably longer and has a timescale of 14
days. This is expected only for a very small fraction of self-lensing
events \citep[compare the density contours
of][Fig. 9]{2006ApJS..163..225R} and is more easy to reconcile with
halo lensing.

The events PAnd-5 and PAnd-6 finally are at $19.3'$ and $27.8'$ or 4.5 and
6 kpc from the M31 center along the major axis.  This is the location
where the SFB from the bulge drops to or below that of the young disk
component, which is about 4 magnitudes below the SFB of the disk. At
this distance from the M31 center bulge lensing should be almost zero
and disk-disk self-lensing should be (because of the inefficient
geometry due to the small line of sight extension) small as well.
Halo lensing by MACHOs will decrease less fast out to 6 kpc since the
disk stars provide a spatially almost `flat' (compared to the bulge)
resource for halo lensing at 2-6 kpc. The timescale of the two outer
lensing events is 3 (PAnd-6) and 9 (PAnd-5) days. The colors are both
0.7 mag (in either $\rps-\ips$ or $R-I$), which is in agreement both with evolved bulge and disk stars.
Since PAnd-5 is at a location where dust lanes become visible we also
checked the extinction values along the line of sights to the events,
to find out, whether one event could be particularly reddened by M31
dust. We used the dust extinction map of \cite{2009A&A...507..283M}
(the spatial resolution of the dust map is $6''$/pixel) and
obtained E(B-V) values of 0.17 to 0.2 for all 6 events. According to
this map PAnd-5 is close but not along the line of sight to a larger
dust extinction.

Our results concerning these long timescale events at 2 and 6 kpc from
the Andromeda center are in agreement with previous results: Long time
scale lensing events have been found by POINT-AGAPE
\citep{2005A&A...443..911C} at a distance of $22'$ from M31 (PA-99-N2)
with a times scale of 22 days.

We also inspect the HST images at the location of these events (see
Fig. \ref{fig.GL_HST}).  At this point we assume that the world
coordinate system (WCS) of the HST images is correct and we do not map
them to the PS1 images.  There are HST archive images for all the
events except PAnd-5. At the position of PAnd-3, PAnd-4 and PAnd-6,
there are one to several resolved sources around the position of our
candidates, while for PAnd-2 there is a possible source and for PAnd-1
nearby sources can be ruled out.

We have not found extended sources (background galaxies) overlapping
or close to our events, so we can rule out the contamination from
SNe. This is also confirmed by the good $\chi _{\mathrm{dof}}$ from
fitting the microlensing light curve.

\begin{figure}[!h]
  \centering
  \includegraphics[scale=0.8]{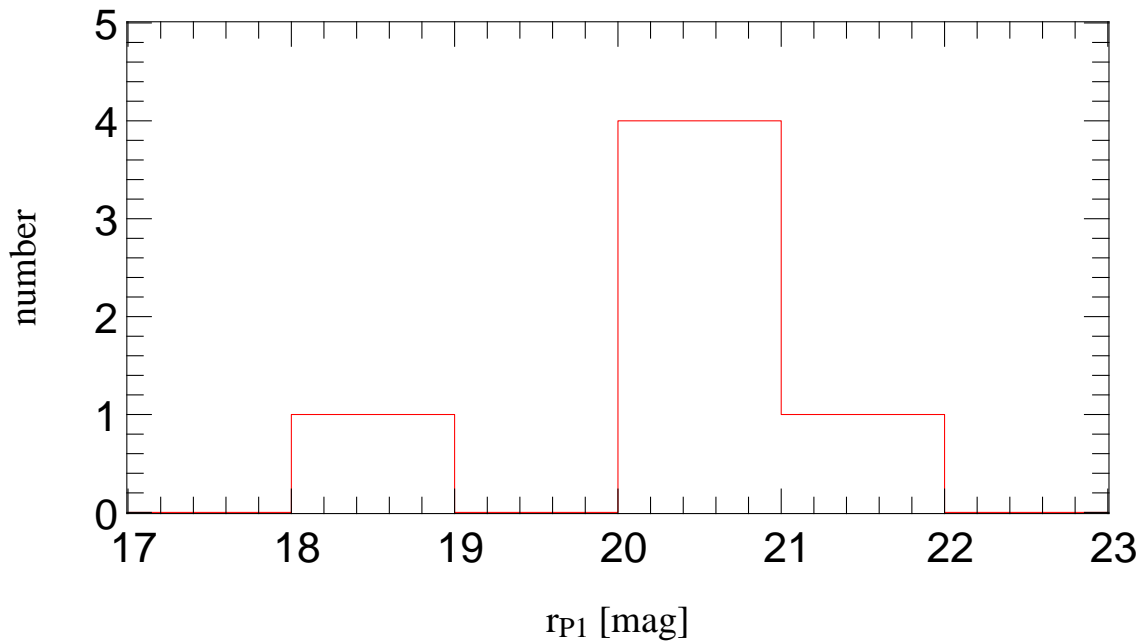}
  \caption{Distribution of the $F _\mathrm{eff,\rps}$ of the 6 PAndromeda
microlensing events from Table \ref{tab.ML}.  
Most of the events have $\rps$ fainter than 20
mag at the flux maximum except PAnd-3. Such a bright event is hard to
reconcile with self-lensing due to the flux excess limit caused by
finite-source effects.  In general very bright events can be more
easily caused by MACHOs \citep{2008ApJ...684.1093R}.}
  \label{fig.mag_histo}
\end{figure}

\begin{figure}[!h]
  \centering
  \includegraphics[scale=0.8]{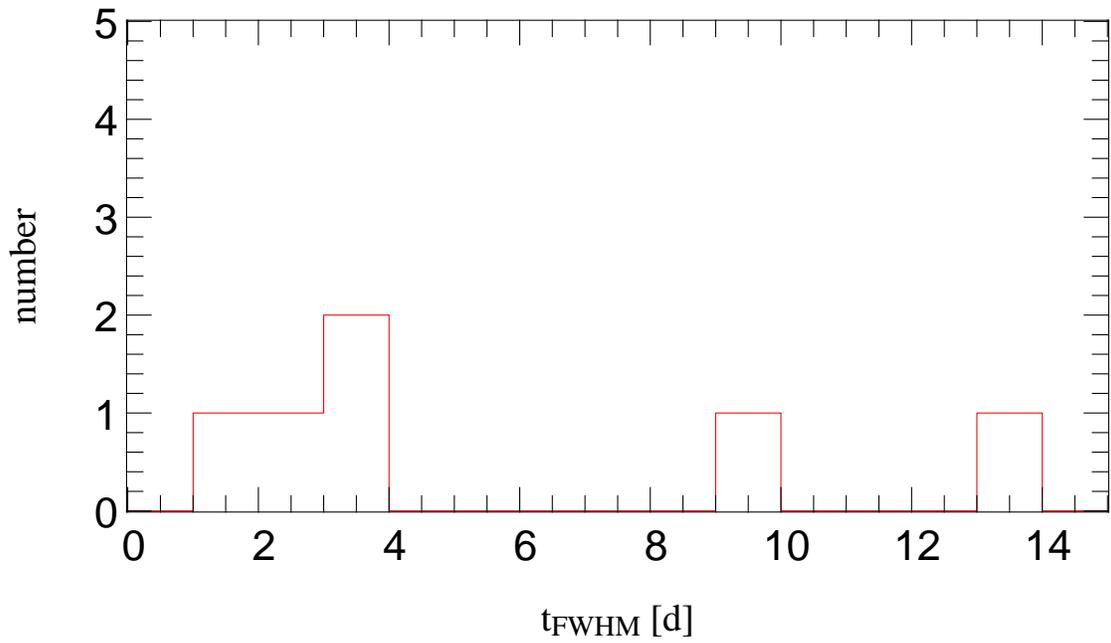}
  \caption{Distribution of the $\tfwhm$ of the 6 PAndromeda
microlensing events.  The timescale of the detected events ranges from
1.3 to 13.8 days, with 4 out of the 6 events have $\tfwhm$ below 5
days.  This demonstrates that the PAndromeda survey is effective in
finding short duration events.}
  \label{fig.tfwhm_histo}
\end{figure}

\begin{figure*}
  \centering
  \includegraphics[angle=270,scale=0.35]{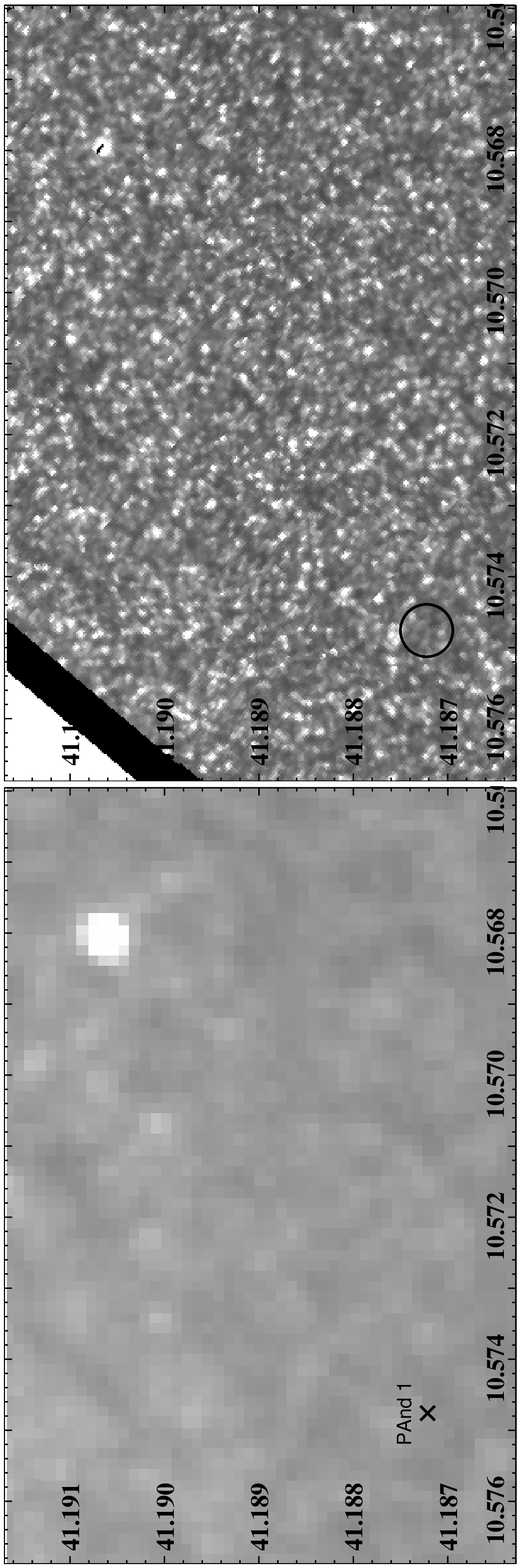}
  \includegraphics[angle=270,scale=0.35]{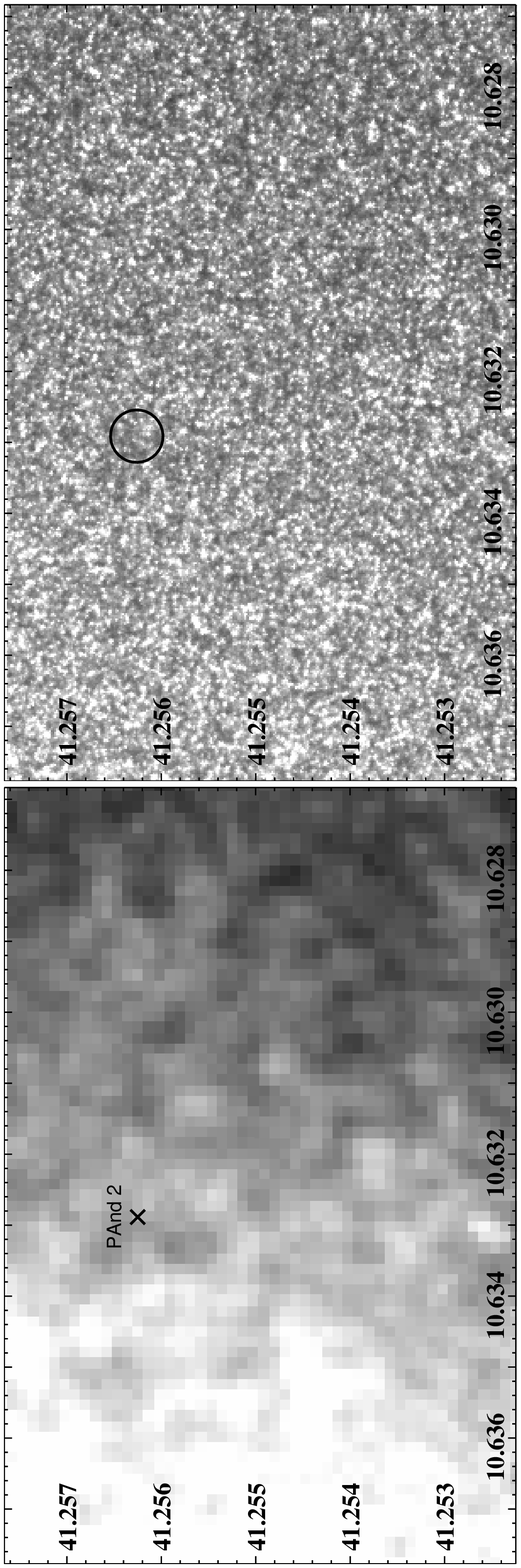}
  \includegraphics[angle=270,scale=0.35]{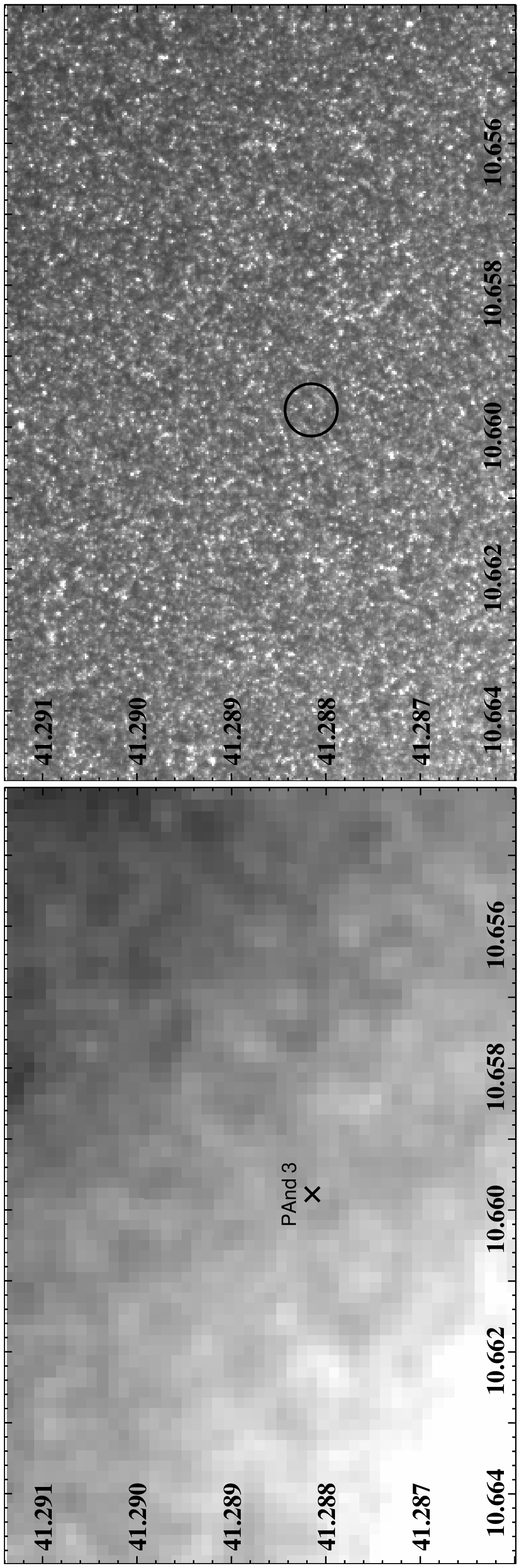}
  \includegraphics[angle=270,scale=0.35]{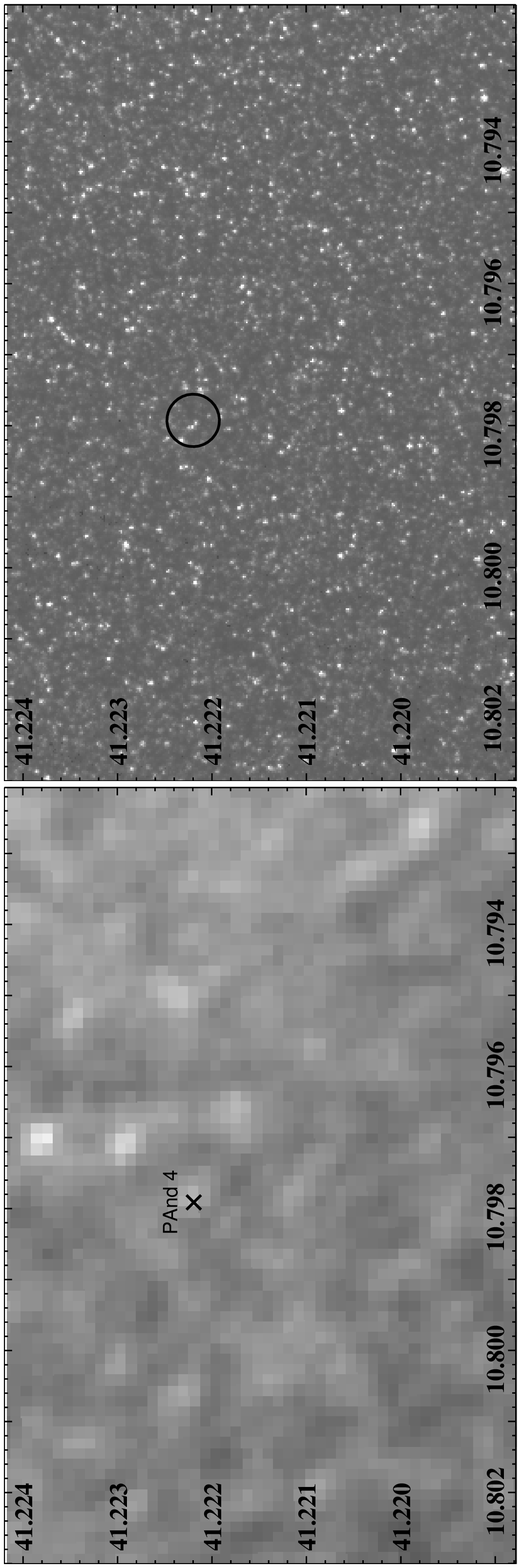}
  \includegraphics[angle=270,scale=0.35]{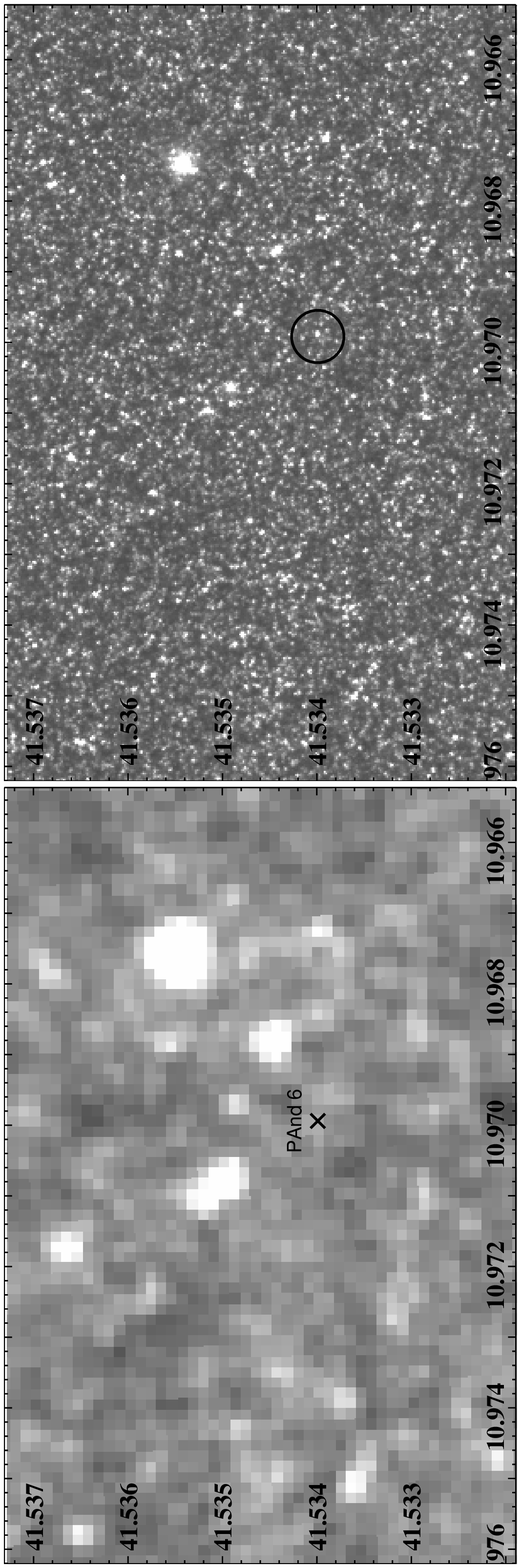}
  \caption{Comparison of PS1 images (left column) with the HST images
(right column).  The HST image of PAnd-1 is from a median of 3 ACS 814
nm archive data (u3d9020at\_drz.fits, u3d9020gt\_drz.fits,
u3d9020mt\_drz.fits) of HST Cycle 5 proposal 6300 by Ford, H.. 
The HST image of PAnd-2,
PAnd-3 and PAnd-4 are from ACS 814 nm archive data
(jbf309010\_drz.fits, jbf308010\_drz.fits and jbf310010\_drz.fits) of
HST Cycle 18 proposal 12058 by Delcanton, J.. The HST image of PANd-6 is
from ACS 606 nm archive data (j96q05010\_drz.fits) of HST Cycle 13
proposal 10407 by Morrison, H..}
  \label{fig.GL_HST}
\end{figure*}

\clearpage

\section{Outlook}
\label{sec.outlook}

We are currently analyzing the full data set of the first PAndromeda
season. Besides the point-source point-lens modeling and detection
presented in section \ref{sec.ML}, a search for the finite-source
effects according to the process by \cite{2009ApJ...695..200L} will
also be carried out. Further investigations on the individual bright
events following the analysis of \cite{2008ApJ...684.1093R} will shed
light on the identity of these events, whether they are more likely
caused by MACHOs than by stellar lenses.  A comprehensive study of the
detection efficiency (e.g. as a function of $\Delta_{F}$ and the event
duration in the different fields) will also be done.

In addition to the microlensing analysis, we will use the PAndromeda
data-set to improve the M31 model in several aspects, for example,
including the dust content \citep{2009A&A...507..283M}, modeling the
three-dimensional distribution of stars. At the time of writing this
paper, we are also granted time to investigate the stellar population
and dynamics of the M31 bulge in more detail (10 nights of the 2.7m
telescope at McDonald observatory, with the VIRUS-W instrument),
e.g. the bulge-bulge self-lensing (main contribution to the
self-lensing in the center of M31) can be quantified at a much better
level than now.

The IPP is going to deliver a reprocessing of the
PAndromeda data in some months from now.  We will analyze the
reprocessed data and also the upcoming new data without binning.  The
observation strategy in 2011 is similar to 2010. We have two rotation
angles of the de-rotator both with a good matching on a grid of
corresponding detectors.  In 2010 the stars on the central part of the
cameras were defocused, leading to a decreased S/N.  This can mimic an
asymmetric microlensing signal.  Therefore for 2011 we shift this
defocused center to the major axis of M31 to have more symmetric
signals from both sides of M31 (see Fig. \ref{fig.gpc}). Of course,
our final detection efficiency study will have to account for the
varying PSF sizes in different skycells.

In 2011 we continue with our two block strategy with a time gap of
three to five hours accepting short offsets of $<$30 min due to
scheduling constraints.  The filters and exposure times will be the
same like in the previous season.

The pointing follows a dither pattern with a diameter of $13''$ to $20''$
and a cycle of 5 images.  This ensures that all the detector gaps are
filled completely.  The regular observations have started on 25th July
2011.

\section{Conclusion}
\label{sec.conclusion}
The preliminary analysis of the first part of the first season
PAndromeda data demonstrates that one can indeed detect microlensing
events at a competitive rate per season \citep[see][for a summary of
previous surveys and events]{2010GReGr..42.2101C}.  The data are
further useful for novae detection and investigation of other
variables.

The identification of 4 short-duration microlensing events with
$\tfwhm \sim$ 1-3 days shows that the time resolution of the
PAndromeda project is comparable with the best 2 seasons of the WeCAPP
project (where two telescopes were coordinated to monitor M31). The
advantage of PAndromeda is the capability to cover the entire M31 disk
with one pointing (compared to the $17.2' \times 17.2'$ FOV of WeCAPP).
By finding or excluding microlensing events in the outer disk of M31,
the PAndromeda survey will manifestly exceed the accuracy of
previously derived M31-halo MACHO fractions.

\noindent
\textit{Acknowledgments}
\indent

We thank for comments from the anonymous referee. We are grateful to 
Sebastiano Calchi Novati and Jelte de Jong for their useful comments.
This work was supported by the DFG cluster of excellence `Origin and
Structure of the Universe' (www.universe-cluster.de).

The Pan-STARRS1 Survey has been made possible through contributions of
the Institute for Astronomy, the University of Hawaii, the Pan-STARRS
Project Office, the Max-Planck Society and its participating
institutes, the Max Planck Institute for Astronomy, Heidelberg and the
Max Planck Institute for Extraterrestrial Physics, Garching, The Johns
Hopkins University, Durham University, the University of Edinburgh,
Queen's University Belfast, the Harvard-Smithsonian Center for
Astrophysics, and the Las Cumbres Observatory Global Telescope
Network, Incorporated, the National Central University of Taiwan, and
the National Aeronautics and Space Administration under Grant
No. NNX08AR22G issued through the Planetary Science Division of the
NASA Science Mission Directorate.

Funding for the SDSS and SDSS-II has been provided by the Alfred
P. Sloan Foundation, the Participating Institutions, the National
Science Foundation, the U.S. Department of Energy, the National
Aeronautics and Space Administration, the Japanese Monbukagakusho, the
Max Planck Society, and the Higher Education Funding Council for
England. The SDSS Web Site is http://www.sdss.org/.

The SDSS is managed by the Astrophysical Research Consortium for the
Participating Institutions. The Participating Institutions are the
American Museum of Natural History, Astrophysical Institute Potsdam,
University of Basel, University of Cambridge, Case Western Reserve
University, University of Chicago, Drexel University, Fermilab, the
Institute for Advanced Study, the Japan Participation Group, Johns
Hopkins University, the Joint Institute for Nuclear Astrophysics, the
Kavli Institute for Particle Astrophysics and Cosmology, the Korean
Scientist Group, the Chinese Academy of Sciences (LAMOST), Los Alamos
National Laboratory, the Max-Planck-Institute for Astronomy (MPIA),
the Max-Planck-Institute for Astrophysics (MPA), New Mexico State
University, Ohio State University, University of Pittsburgh,
University of Portsmouth, Princeton University, the United States
Naval Observatory, and the University of Washington.

\bibliographystyle{aamod}
\bibliography{literature}

\end{document}